\documentclass[12pt]{article}

\usepackage{amssymb,epsf,amsbsy}

\def\lsim{\mathrel{\rlap {\raise.5ex\hbox{$ < $}}

{\lower.5ex\hbox{$\sim$}}}}

\def\gsim{\mathrel{\rlap {\raise.5ex\hbox{$ > $}}

{\lower.5ex\hbox{$\sim$}}}}

\def\sqr#1#2{{\vcenter{\vbox{\hrule height.#2pt

        \hbox{\vrule width.#2pt height#1pt \kern#1pt

           \vrule width.#2pt}

        \hrule height.#2pt}}}}

\def\lsim{{\displaystyle
{{\raise-8pt\hbox{$ <$}}
\atop{\raise5pt\hbox{$\sim$}}}}}
\def\gsim{{\displaystyle
{{\raise-8pt\hbox{$ >$}}
\atop{\raise5pt\hbox{$\sim$}}}}}
\def\slsim{{\displaystyle
{{\raise-8pt\hbox{$\scriptstyle <$}}
\atop{\raise5pt\hbox{$\scriptstyle \sim$}}}}}
\def\sgsim{{\displaystyle
{{\raise-8pt\hbox{$\scriptstyle  >$}}
\atop{\raise5pt\hbox{$\scriptstyle \sim$}}}}}
\newskip\humongous \humongous=0pt plus 1000pt minus 1000pt

\newcommand{\sumpf}[0]{\sum_{(H^{\rm f},G^{\rm f})}\! \! \! \!
{\raise
4pt
\hbox{$'$}}\,}

\newcommand{\sump}[0]{\sum_{(H,G)}\! \! {\raise 4pt \hbox{$'$}}\,}

\def\bs{\begin{subequations}}
\def\es{\end{subequations}}
\catcode`\@=11
\newcount\hour
\newcount\minute
\newtoks\amorpm
\hour=\time\divide\hour by60
\minute=\time{\multiply\hour by60 \global\advance\minute by-\hour}
\edef\standardtime{{\ifnum\hour<12 \global\amorpm={am}%
        \else\global\amorpm={pm}\advance\hour by-12 \fi
        \ifnum\hour=0 \hour=12 \fi
        \number\hour:\ifnum\minute<10 0\fi\number\minute\the\amorpm}}
\edef\militarytime{\number\hour:\ifnum\minute<10 0\fi\number\minute}
\def\draftlabel#1{{\@bsphack\if@filesw {\let\thepage\relax
   \xdef\@gtempa{\write\@auxout{\string
      \newlabel{#1}{{\@currentlabel}{\thepage}}}}}\@gtempa
   \if@nobreak \ifvmode\nobreak\fi\fi\fi\@esphack}
        \gdef\@eqnlabel{#1}}
\def\@eqnlabel{}
\def\@vacuum{}
\def\draftmarginnote#1{\marginpar{\raggedright\scriptsize\tt#1}}
\def\draft{\oddsidemargin -.2truein
        \def\@oddfoot{\sl preliminary draft \hfil
        \rm\thepage\hfil\sl\today\quad\militarytime}
        \let\@evenfoot\@oddfoot \overfullrule 3pt
        \let\label=\draftlabel
        \let\marginnote=\draftmarginnote
   \def\@eqnnum{(\theequation)\rlap{\kern\marginparsep\tt\@eqnlabel}%
\global\let\@eqnlabel\@vacuum}  }
\def\subequations{\refstepcounter{equation}%
  \edef\@savedequation{\the\c@equation}%
  \@stequation=\expandafter{\theequation}%   %only want \theequation
  \edef\@savedtheequation{\the\@stequation}% % expanded once
  \edef\oldtheequation{\theequation}%
  \setcounter{equation}{0}%
  \def\theequation{\oldtheequation\alph{equation}}}

\def\endsubequations{\setcounter{equation}{\@savedequation}%
  \@stequation=\expandafter{\@savedtheequation}%
  \edef\theequation{\the\@stequation}\global\@ignoretrue
  \vspace*{-12pt} \\}
\def\bs{\begin{subequations}}
\def\es{\end{subequations}}
\relax
\def\tr{\,{\rm tr}\, }
\def\Im{\,{\rm Im}\, }

\def\thefootnote{\fnsymbol{footnote}}
\def\be{\begin{equation}}
\def\ee{\end{equation}}
\def\ba{\begin{eqnarray}}
\def\ea{\end{eqnarray}}

\def\ee{\end{equation}}
\def\bea{\begin{eqnarray}}
\def\eea{\end{eqnarray}}
\def\nn{\nonumber}

\newcommand{\uarrw}[0]{\mathrel{
{\raise.5ex\vbox{\hrule width 1cm}\hskip-6pt\rightarrow}}}
\def\thebibliography#1{%
\vskip 0.5cm \centerline{\bf References}
\list{%
[\arabic{enumi}]}{\settowidth\labelwidth{[#1]}
\leftmargin\labelwidth
\advance\leftmargin\labelsep
\usecounter{enumi}}
\def\newblock{\hskip .11em plus .33em minus .07em}
\sloppy\clubpenalty4000\widowpenalty4000
\sfcode`\.=1000\relax}

\renewcommand{\theequation}{\arabic{section}.\arabic{equation}}

\renewcommand{\section}{\setcounter{equation}{0}\@startsection%
{section}{1}{0mm}{-\baselineskip}{0.5\baselineskip}%
{\normalfont\normalsize\bfseries}}

\renewcommand{\subsection}{\@startsection%
{subsection}{2}{0mm}{-\baselineskip}{0.5\baselineskip}%
{\normalfont\normalsize\slshape}}

\renewcommand{\subsubsection}{\@startsection%
{subsubsection}{2}{0mm}{-\baselineskip}{0.5\baselineskip}%
{\normalfont\normalsize\slshape}}

\begin{document}
%
%\special{!userdict begin /bop-hook{gsave 200 30 translate
%65 rotate /Times-Roman findfont 216 scalefont setfont
%0 0 moveto 0.85 setgray (\jobname) show grestore}def end}
% 
\renewcommand{\theequation}{\arabic{section}.\arabic{equation}}
\begin{titlepage}
\begin{flushright}
HU-EP 01/46,\\
hep-th/0110201 
\end{flushright}
\begin{centering}
\vspace{1.0in}
\boldmath
{\bf \large String--String triality for ${\bf d=4}$, 
${\bf Z_2}$ orbifolds$^\dagger$}
\\
\unboldmath
%\vspace{.15in}
\vspace{1.7 cm}
{\bf Andrea Gregori} \\
\medskip
\vspace{.4in}
{\it  Humboldt-Universit\"at, Institut f\"ur Physik}\\
{\it D-10115 Berlin, Germany}\\

\vspace{3.2cm}
{\bf Abstract}\\
\vspace{.2in}
\end{centering}
We investigate the perturbative and non-perturbative correspondence
of a class of four dimensional dual string constructions with ${\cal N}=4$
and ${\cal N}=2$ supersymmetry, obtained as $Z_2$ or $Z_2 \times Z_2$
orbifolds of the type~II, heterotic and type~I string. In particular,
we discuss the heterotic and type~I dual of all the symmetric
$Z_2 \times Z_2$ orbifolds of the type~II string, classified in Ref.
\cite{gkr}. 

\vspace{4cm}

\hrule width 6.7cm
\noindent
$^\dag$\  Research supported by the ``Marie Curie'' fellowship
HPMF-CT-1999-00396.\\
$^1$e-mail: agregori@physik.hu-berlin.de

\end{titlepage}
\newpage
\setcounter{footnote}{0}
\renewcommand{\thefootnote}{\arabic{footnote}}

\tableofcontents

\vspace{1.5cm}

\noindent

\section{\bf Introduction}

After the appearance of Refs. \cite{w} and \cite{pw}, 
the duality between the heterotic, the type~II and the type~I string in less
than six dimensions, and notably in four dimensions, has been one
of the major subjects of investigation some years ago.
In some specific cases, tests of this duality have been provided 
\cite{kv,klm}, 
and, once understood some general ``rules'' for the identification of string 
duals \cite{al}, started a ``systematic'' investigation of certain 
non-perturbative properties of string constructions, in particular
of the heterotic string \cite{kklm,mvI,mvII,wsi,am,aspin} \footnote{For a 
review of more recent results, see \cite{mayrrev}.}.
However, most of these analyses, although applied to string theory,
as a matter of fact rely more on supergravity, field theory and
the properties of supersymmetry or even of algebraic geometry, 
than on real properties of the string theory as such, deeply distinguished from
any of the legitimate ingredients of its effective descriptions.
In fact,  the relation between string theory, namely
a theory expected to be ``the'' quantum theory of gravity and gauge 
interactions, and its representations in terms of supergravity or 
field theory, allowed in certain limits, is rather subtle. 
As a matter of fact, many tests
of string--string duality are actually no much more than
consistency tests of the duality among the underlying effective 
supergravities. What makes a hard task to really test string--string dualities
in a string framework, besides the lack of knowledge of a 
non-perturbative definition of the theory, is that, already at the
perturbative level, the partition function at a generic point of 
the perturbative moduli space is not known. We nevertheless consider such tests the only way 
of getting, when possible, reliable informations, 
minimizing thereby the possibility of falling into the trap of  
``tautologies'', a risk always present when investigating string dualities
through the projection onto a ``sub-sector'' of the theory, such as 
supergravity or field theory.  

For similar reasons, we also still consider direct string constructions 
something based on a more solid ground than string vacua
obtained through the reduction of higher dimensional theories such as,
for instance, F-theory. 
The only way of testing the reliability of these vacua is through
the comparison with string theory; once these conditions have been
taken into account, these vacua turn out to provide no more informations
than what one can already obtain from the string theory itself.

In this work we therefore investigate string--string duality
in four dimensions by taking a ``conservative'' attitude,
going back to ``simple'' examples for which we have
direct, explicit string constructions. 
Our aim is to provide an analysis free of the prejudices 
that may be induced by supergravity/field-theory considerations, 
not always appropriate to the analysis of superstrings. 
String--string duality is in fact more fundamental
than supergravity dualities, and has its origin in the expected
equivalence of the various ``regularization'' procedures
of the ultraviolet divergences, corresponding to the different 
``types'' of strings. The duality between string constructions therefore
does not necessarily imply a duality of the corresponding 
effective supergravity theories \footnote{This is particularly true when
in some constructions part of the spectrum is non-perturbative.
Although they do not appear explicitly in the spectrum, and therefore
neither in the associated supergravity, string theory nevertheless
``knows'' about the ``missing'' states, to which supergravity is
instead totally ``blind''.}. Taking this into account, we 
decided to base our investigation on \emph{direct string computations}, 
extracting as much information as possible from the comparison of dual string 
constructions,  and avoiding any non-necessary assumption.
In practice, we reconsider the heterotic/type~II/type~I string--string duality
in the case of ${\cal N}=4,2$ supersymmetry in four dimensions, 
for $Z_2$ orbifold constructions \footnote{The convenience of working 
with $Z_2$ orbifolds is that in this case not
only we explicitly know the partition function but it is also easy to
compute the moduli dependence of certain amplitudes.}.
As is known, in less than ten dimensions it may happen that not all
the massless degrees of freedom appear explicitly in the perturbative spectrum
of a certain string construction, even at weak coupling.
This is the case of certain gauge states for the type~II string,
in which the states charged under the RR fields 
don't have a representation in terms of vertex operators, and do not appear
in the perturbative spectrum; or the case of massless states appearing when 
heterotic instantons shrink to zero size.  
This makes the identification of string dual pairs something that
has to be handled with care, starting from the very choice of the
criteria that allow to identify such pairs: we in fact cannot expect
to always have a perturbative--perturbative duality, with an
exact correspondence of the two perturbative effective theories. 
This leaves open an apparent arbitrariness in the identification of 
dual pairs. For instance,
the fact that, although existing, gauge charged states never appear on the 
type~II side, implies that a wide choice of gauge configurations
on the heterotic side may correspond to the \emph{same} type~II construction.
The case of heterotic ``small instantons'' is particularly delicate:
such states may appear in fact for special values of the moduli
in the hypermultiplets, and the $T^4 \big/ Z_2$ orbifold point of the K3
is precisely a special point in the hypermultiplets space.
We should therefore not be too surprised to find that, in some cases,
precisely at this point the massless spectrum possesses a non-perturbative 
extension. Although deceiving this may look, it has not to be taken as a total
impossibility of determining dual pairs.
The point of view we took is that one should not expect a naive coincidence of
massless spectra, often forbidden by accidental technical facts, such as the
impossibility of constructing certain vertex operators, nor an automatic
coincidence of the moduli spaces of the supergravity theories associated to
the perturbative spectrum, or, even worse, a ``fiber-wise'' correspondence
of the geometric spaces on which the various ten dimensional strings
are compactified: a geometric correspondence doesn't imply in general
a correspondence of the quantum moduli spaces \footnote{Remember that
dual string constructions can be compared only in the Einstein frame,
not in the string frame!}. Rather, our general guidelines have been:
\newline
\romannumeral1) to first identify the general criteria of why certain states 
may or may not appear in the perturbative spectrum of a certain string 
construction, and only after the identification of which subsets of
the massless spectrum of the string theory may appear in the
specific, heterotic or type~II, or type~I construction,
to proceed to the comparison of the spectra
of different string constructions.
\newline
\romannumeral2) to rely on the comparison of quantities explicitly,
independently computed on the various string constructions,
avoiding therefore any use of ``index'' theorems, or conjectures based
on the properties of supersymmetry. We did not want in fact
to exclude a priori the possibility that exact supersymmetry may be only
an approximation.
\newline
\romannumeral3) to require compatibility with the perturbative properties
(spectrum, symmetries) of the various string realizations, 
in the appropriate weak coupling limits.

Even by proceeding in this way, a certain amount
of arbitrariness and assumptions has been necessary. This is
somehow unavoidable; otherwise, any duality test would be
``tautological'', and string-string duality would loose
its major interest, namely the predictive power.
Nevertheless, we find that the criteria we used are quite reasonable
and, as it will be clear from the analysis of the various cases,
very restrictive.
Although we cannot ``prove'' the dualities, it is 
encouraging that the picture we find is extremely consistent,
and allows to clarify several aspects of string theory.

\vspace{0.9cm}

The paper is organized as follows:

After a brief introduction to the subject in section \ref{hIIAI},
our analysis begins with a review of string--string duality 
for ${\cal N}=4$ supersymmetry, in section \ref{z2orbifolds}. 
This section serves us to clarify certain points, that will reappear
in the subsequent sections, and to start ``setting the rules'' of the 
analysis. One of the quantities that will guide the following investigation is
the mass formula of the string states that are ``lifted'' by freely
acting projections. In such cases, the masses of the states acquire
a dependence on the string moduli. Depending on whether these moduli
are perturbative or non-perturbative, and whether the masses are
suppressed or blow up in the perturbative limit, these states may or may
not appear in a certain string construction. This is therefore an
essential key in order to properly identify the string duals: in particular,
it was the guideline for a new orientifold construction, presented in section
\ref{sbroken}, in which the world-sheet parity projection acts freely.

We pass then, in section \ref{n=2}, to the analysis of ${\cal N}=2$
orbifolds, freely and non-freely acting.
In the first case, the massless spectrum is just a projection of the
dimensionally reduced supergravity spectrum of ten dimensions. 
In the second case the spectrum contains additional states, associated to the
orbifold fixed points, and originating from the twisted sectors.
On the heterotic side, these states are hypermultiplets,
while on the type~IIA and type~I side contain in general also new vector 
multiplets. On the type~I side, the ``twisted'' sector appears in the form
of a D5 branes sector. The case
of non-freely acting orbifolds proves to be the most interesting one:
precisely  in this case, on the type~IIA~side the compact space cannot be 
interpreted as a limit of a K3 fibration. Therefore, there cannot be  a
``perturbative'' correspondence with the heterotic string: the
type~IIA spectrum contains in fact states that are of non-perturbative origin
from the heterotic point of view. This however by no means forbids
a test of string--string duality. These states are in fact of the 
``small instantons'' type; they are therefore present even in the heterotic 
weak coupling limit, they interact with the ``perturbative'' ones, and 
indirectly contribute to the renormalization
of quantities that we can compute in string perturbation.
Quite remarkably, it is possible to define amplitudes, 
built on $R^2$ and $F^2$ terms, that allow to unambiguously test
string--string duality through a comparison of direct computations 
performed on the type~II, heterotic and type~I sides \cite{gkr,gkp,gkp2,gk}. 
As we will see, when the dual 
constructions are properly identified, there is an astonishing
correspondence of these amplitudes: the contributions
of the moduli related by duality appear through the same function,
with a coincidence even in its normalization.

The analysis of duality in the case of freely acting, or semi-freely acting, 
orbifolds, makes on the other hand use, as we said, of the informations 
provided by the moduli-dependence of the mass of certain states.
Particularly interesting are the cases of those semi-freely acting orbifolds
in which the free part of the orbifold action is visible only on the
type~II, or type~I side, because it involves an action on the modulus
dual to the heterotic dilaton. From the heterotic point of view, they appear
therefore as ordinary, non-freely acting orbifolds, from which they
are distinguished only by the non-perturbative behavior, at non-zero
string coupling. Again, the (heterotic) perturbative physics is not 
however completely blind to these phenomena, and still the heterotic dual 
can be unambiguously identified. From string--string duality,
we then learn that, from the heterotic point of view,
in these constructions the gauge group is of entirely non-perturbative origin.

The identification of the mechanism of rank reduction, in all the
dual string constructions, allows then us to complete the analysis of the 
${\cal N}_4=2$, $Z_2$ orbifolds, and ultimately provide the heterotic and 
type~I dual for all
the $Z_2 \times Z_2$ type~II orbifolds of Ref. \cite{gkr}, i.e. of 
all the $Z_2 \times Z_2$ orbifolds of the type~II string in which
each $Z_2$ projection acts as a left--right symmetric twist of four
coordinates.

All the string vacua considered in section \ref{n=2}
present gauge groups with vanishing (one loop) gauge beta-function.
In ${\cal N}_4=2$ string theory a  non-vanishing gauge beta-function always 
introduces in the expression of the effective coupling other moduli, among
which there are those parameterizing the coupling constant of ``hidden'' 
gauge sectors, of the ``small instantons'' type.  
The appearance of these couplings raises a puzzle of ${\cal N}_4=2$
string theory. On one hand, we expect in fact
the string corrections to the gauge coupling, as it is sensitive to
the couplings, to be also sensitive to the mass
of the states of these hidden sectors, in such a way that, 
as they become massive and decouple,
their contribution to the correction vanishes; on the other hand,
the masses of these states depend on the moduli in the hypermultiplets,
and these are forbidden by ${\cal N}_4=2$ supersymmetry to enter
in the expressions of the corrections to the gauge couplings. 
In section \ref{sbreaking} we propose  a physical interpretation
that allows to solve this paradox.
By considering the peculiar way the couplings of the hidden
sectors enter into the gauge corrections, we argue that, as soon as 
there is a non-vanishing gauge beta-function, the visible sector
interacts with the hidden sector, which is now in a phase of strong coupling.
Supersymmetry is therefore broken because of gaugino condensation
in the hidden sector. The breaking of supersymmetry is non-perturbative
from the point of view of the visible sector, that therefore 
appears, in a perturbative construction, as an expansion around
an ${\cal N}_4=2$ vacuum. However, the fact that supersymmetry is broken
forbids the use of ``index'' theorems in order to promote the
string corrections, as explicitly computed for instance in the heterotic 
or type~I string at the orbifold point, 
to a result valid at other points in the moduli of the compact space.
According to this interpretation, results such as those of
Refs. \cite{hm,agn1,agn2,ms} would be
universally valid only in the case of vanishing gauge beta-functions.

The breaking of supersymmetry in theories with non-vanishing gauge 
beta-functions is peculiar of string theory, and does not have
a counterpart in field theory, in which a beta-function does not
imply an interaction with ``hidden'' sectors.

\vspace{1.5cm}

\noindent

\section{\bf Heterotic/type IIA/type I duality}
\label{hIIAI}

In this section we review some old topics about string-string
duality between the heterotic, the type~IIA and the type~I string
in (six and) four dimensions. Throughout the paper, we will
use the convention to indicate the number ${\cal N}$ of space-time
supersymmetries in $d$ dimensions as ${\cal N}_d$.

\subsection{\sl Heterotic/type IIA duality }

We start by considering the duality between the heterotic and the type~IIA
string. We consider here general properties,
valid for compactifications on spaces more general than the $Z_2$
orbifolds, to which we will restrict from section \ref{z2orbifolds}.

\subsubsection{\it ${\cal N}_6=2$ and ${\cal N}_4=4$}

The heterotic string toroidally
compactified to six dimensions has been conjectured to be
dual to the type~IIA string
compactified on a K3 \cite{w,dk,ht}, through an inversion of coupling. 
When further compactified on a 
two-torus, the two four-dimensional theories are then mapped the one into
the other by exchanging a modulus of the gravity manifold (the 
dilaton--axion field $S$) with a modulus, $T$, of the vector manifold 
\cite{dk,ht}. 
This conjecture of duality is based on the comparison of the low
energy effective actions of the two theories, and is supported
by several tests, essentially  relying on properties typical of the extended
supersymmetry \cite{kv,klm,al}.
The reason why tests based on extended supersymmetry properties are 
very reliable is that in the ${\cal N}_6=2 \big/ {\cal N}_4=4$ case 
all the non-perturbative phenomena are suppressed in the weak coupling
limit. The field content of the theory, that determines the
corresponding effective supergravity, is easily accessible
via a perturbative analysis on the heterotic side, and can be compared
with what one derives from a geometric analysis on the type~IIA side,
``simple'' because of the uniqueness of the K3 surface.
Once identified the underlying supergravity/super Yang--Mills theory,
one then simply uses the properties of extended supersymmetry.

\subsubsection{\it ${\cal N}_4=2$}
\label{n2}

More complicated is the case in which supersymmetry is reduced by 
half (in this case, the heterotic/type~IIA duality exists only in four 
dimensions). On the heterotic side, such a reduction of supersymmetry 
is obtained, in the case of constructions that admit a geometric
interpretation, by compactifying six coordinates on $T^2 \times K3$
and choosing an appropriate gauge bundle. The K3 surface admits
several orbifold limits. In these cases, the orbifold operation in four
of these six coordinates can be coupled to translations acting on $T^2$. 
An analysis based on the properties of the 
effective ${\cal N}=2$ supergravity theory, as derived from the spectrum of
the heterotic string, indicates that the type~IIA dual of 
such a heterotic construction has to be found among the type~IIA
vacua obtained by compactification on a Calabi--Yau manifold 
which is also a K3 fibration \cite{al}. This observation is based on 
the analysis of the prepotential, a holomorphic function of the
moduli of the vector manifold, that encodes the informations 
about the couplings involving the vector multiplets. 
The prepotential corresponding to the effective theory
built on the perturbative massless fields of the
heterotic compactification reads:
\be
{\cal F} = S( TU - \sum_i C^i C^i ) + {\rm F}_0 ( T, U, C^i)+
{\rm F}_{\rm n.p.} ( {\rm e}^{- S}, T, U, C^i) \, , 
\label{prep}
\ee
where $T$, $U$ are the moduli associated 
respectively to the K\"{a}hler class and the complex structure of the 
two-torus, while $C^i$ are associated to the scalars in the vector
multiplets originating from the currents. The term
${\rm F}_0 ( T, U, C^i)$ contains terms of the type
$c_{abc}y^a y^b y^c$, cubic in these fields.  
The heterotic dilaton--axion field $S$ corresponds to the
complex modulus associated to the volume of the base of the fibration.

For particular configurations
of the K3 space, the heterotic string contains additional vector multiplets, 
entirely non-perturbative. These states appear when
instantons shrink to zero size. This phenomenon
depends on the particular configuration 
of the K3, i.e. on the moduli in the hypermultiplets, and, although 
of non-perturbative origin, it cannot be eliminated by going to the
very weak coupling limit: this effect in fact is not exponentially suppressed 
by the field $S$. According to the analysis of Ref. \cite{al},
from the dual point of view of the type~IIA string compactified on a K3
fibration, these new states appear when the fiber degenerates.

\noindent

\subsection{\sl Heterotic\big/type I duality in four dimensions}

From the comparison of the effective actions, one can see that
the duality between the heterotic and the
type~I string in four dimensions, in the case of ${\cal N}=4$ 
supersymmetry, is ``perturbative-perturbative'', the heterotic modulus
$S$ being mapped into an analogous modulus of type~I, parameterizing
the coupling of the D9-branes \footnote{The relation between this modulus
and the string parameters is different for the heterotic and the type~I string,
but we will not be concerned with that. Our analysis will always be performed
in the Einstein frame, the one appropriate for the discussion of string-string 
dualities.}.
Duality with ${\cal N}=4$ supersymmetry means that the two theories are 
indeed the same theory. 
However, it is not always easy to identify the duality map
between heterotic and type~I string.    
When supersymmetry is reduced to ${\cal N}=2$, 
in the type~I string there can be additional D-branes sectors,
e.g. D5-branes, and certain states, perturbative on the type~I side,
may map to non-perturbative ones of the heterotic string.
On the other hand, it happens also that certain perturbative moduli of the 
heterotic theory map into non-perturbative ones on the type~I side.
This implies that a perturbative phase of the heterotic string may not 
correspond to a perturbative phase of the type~I string.
An important aspect of the type~I string, that has to be taken into account
when comparing with the heterotic string, is that
there is never mixing between the couplings
of two different sectors that appear explicitly, as D-branes sectors,
on the type~I side. For instance, in the $U(16) \times U(16)$ model
of Ref. \cite{gp}, there are hypermultiplets charged under both the $U(16)$'s.
From a field theory point of view, we would then expect a non-vanishing 
one-loop renormalization of the effective gauge coupling of the
D9-brane sector
depending on the coupling of the D5-branes.
In four dimensions, the first is parameterized by a field $S$, the second 
by a field $S^{\prime}$, and we would expect a correction of the type
\footnote{Our convention for complex fields is $X = X_1 + {\rm i} X_2$.
In particular, the heterotic dilaton--axion field is $S = a + {\rm i} 
{\rm e}^{- \phi}$.}:
\be
{1 \over g^2_9}_{\rm (4-dim)} \, \equiv \, S_2 \; \to \; \approx \, S_2
\, + \, \beta \, S^{\prime \, -1}_2 \, .
\ee
However, on the type~I side this correction is always non-perturbative,
and never appears explicitly. Notice that such a renormalization would exist
even when the D5-branes sector is made massive by some Higgs-mechanism.
Neither the D9-branes nor the D5-branes sector of a type~I model with such a 
structure can therefore be dual to the heterotic currents: 
one of the two couplings would in fact correspond to a perturbative modulus of
the heterotic string, and
its contribution should appear explicitly at the one-loop, 
but with the opposite power (see section \ref{3,51} for more details).
Therefore, even though at the ${\cal N}_4=4$ level the heterotic/type~I duality
implies the identification of the D9-branes sector with the currents
on the heterotic side, when supersymmetry is reduced to  ${\cal N}_4=2$
it is not always true that the D9-branes sector of the type~I 
string maps into the sector corresponding to the perturbative currents of 
the heterotic string.
As we will see, the relation between heterotic and type~I
constructions may be rather non-trivial; in certain cases,
all the D-branes states
correspond to non-perturbative states on the heterotic side.
Our aim in the following is precisely to understand what is the recipe
for the identification of this map, for the moduli in the 
vector multiplets, at least in the simple case of $Z_2$ orbifolds. 

In what follows, we will therefore restrict our analysis to the case of $Z_2$, 
${\cal N}_4=4$ and $Z_2 \times Z_2$, ${\cal N}_4=2$ orbifold constructions of 
the type~II string. We then provide the explicit construction
of the heterotic and type~I duals, for which we also write the partition
function. This will allow us to test our conjectures through direct
\emph{string} computations, thereby going beyond supergravity considerations.

\vspace{1.5cm}

\noindent

\section{\bf ${\bf Z_2}$ orbifolds: ${\bf {\cal N}_4 \boldsymbol{=} 4}$}
\label{z2orbifolds}

In this section we discuss how string-string duality works
in certain examples for which the theory can be realized as a $Z_2$
orbifold of the type~IIA string. From the heterotic (and type~I)
point of view, these theories can be divided into two classes: those with
unbroken and those with broken $SL(2,Z)$ duality of the dilaton--axion field
$S$.

\subsection{\sl unbroken S-duality}
\label{unbroken}

We start by reconsidering the case of ${\cal N}=4$ supersymmetry
in four dimensions. The best known case is the one in which four coordinates
of the type~IIA string are compactified on the $T^4 \big/ Z_2$ orbifold
limit of the K3 surface. The other two are compactified on a torus $T^2$.
At the level of the massless spectrum,
the sixteen fixed points (fixed tori) of this orbifold correspond to
sixteen vector multiplets, dual, on the heterotic side, to the Cartan
subgroup of the gauge group originating from the
currents. In the type~I string, they correspond to massless states
of the D9-branes. Here both the heterotic and type~I string are
compactified on $T^6$. The duality between these constructions has been
widely investigated \footnote{For recent results, see 
Refs.~\cite{6auth,kop}.}, and we will not spend time on this point. In this model,
extended supersymmetry is expected to remain unbroken at any value of 
the coupling (all the gauge $\beta$-functions vanish, there are therefore
no sectors that can be driven to a strong coupling regime, where
supersymmetry is broken  by gaugino condensation 
\footnote{See section \ref{sbreaking}.}).
This allows to make duality tests, by comparing 
terms of the effective action whose renormalization depends on
BPS saturated multiplets of the extended supersymmetry.
In particular, it is useful to consider, as in Refs. \cite{hm4,6auth}, 
the renormalization of the effective
coupling of the so called $R^2$ term.
A calculation performed on the type~IIA side gives:
\be
{16 \, \pi^2  \over g^2} \, = \, -12 \log S_2 | \eta ( \tau_S )|^4 \, ,
\label{seta}
\ee
where the field $\tau_S$ is the K\"{a}hler class modulus of $T^2$,
that we called $\tau_S \equiv 4 \pi S$ because, as is well known, 
is dual to the heterotic field $S^{(\rm het)}$ and to the
type~I field $S^{(\rm type \, I)}$. The large-$S$ behavior of (\ref{seta})
is in fact:
\be
{1 \over g^2} \to S_2 \, ,
\label{larges}
\ee
and matches with the perturbative (``tree level'') expression of 
this coupling, in the heterotic and type~I string.

\subsection{\sl broken S-duality}
\label{sbroken}

Interesting for the following discussion is 
the case in which a twist acting on the currents of the heterotic string makes
massive all the corresponding states.
The explicit type~II construction was presented in Ref. \cite{6auth},
while the heterotic partition function is explicitly written 
in Ref. \cite{gkp2}, for the case in which a further projection
reduces supersymmetry to ${\cal N}_4=2$.
The partition function of the ${\cal N}_4=4$ model is trivially obtained by
setting to zero the action of this second projection in the expressions
of Ref. \cite{gkp2}. This heterotic construction has no
massless states originating from the currents, all of them being lifted
by a set of projections acting freely as twists on the currents and
shifts on the coordinates of the compact space.
The type~IIA dual is realized by coupling the $Z_2$ twist on the four
coordinates to a $Z_2$ translation along a circle of $T^2$, something 
resulting in a shift of the momenta (or the windings, depending on the
choice of the action) corresponding to a certain direction inside $T^2$.
This shift lifts the origin of the lattice;
as a consequence, all the states associated to the previously fixed tori
become now massive. 
In this case, a type~IIA computation of the effective coupling
of the $R^2$ term gives (see Ref. \cite{6auth}, Eq.~(6,6)):
\be
{16 \, \pi^2 \over g^2} = \, -4 \log S_2 | \theta_2 ( \tau_S )|^4 \, ,
\label{stheta}
\ee
whose large-$S$ behavior still matches with what we expect from the heterotic
(and type~I) side, namely expression (\ref{larges}), but 
differs in the fact that now S-duality is broken: the strong coupling
(small-$S$) behavior is very different from the weak coupling
(indeed there is an approximate restoration of the ${\cal N}=8$ 
supersymmetry). Apparently, 
the difference between (\ref{seta}) and (\ref{stheta})
is only in the small-$S$ behavior, something out of the heterotic
perturbative regime: the vector multiplets
associated to the orbifold fixed points have been lifted
by a mechanism that looks purely non-perturbative from the heterotic 
point of view.
It is therefore legitimate to ask \emph{why}
we observe a difference also in the perturbative heterotic string.
With the techniques introduced in Ref.~\cite{kk},
it is possible to follow the fate of the would-have-been massless
states, now lifted. This is obtained by plugging the new momenta and
winding numbers in the BPS mass formula of the ${\cal N}=8$ supersymmetric
type~IIA string, i.e. the theory before the orbifold projection.
We obtain that in first approximation the mass of these states behaves like:
\be
m^2 \sim {1 \over  U_2 \tau_{S_2}}\, .
\label{msu}
\ee
This formula is not exact, because it is derived by applying a shift
on the BPS states of the ${\cal N}=8$ supersymmetry, now broken to
${\cal N}=4$. Most probably, the correct expression is a modular (covariant)
function of these moduli, respecting the symmetries of the
unbroken modular subgroup \footnote{As it happens for 
terms depending on lattice sums,
we expect that here too a mass term in the effective theory
should be given in terms of modular-covariant functions.
Therefore, the meaning of expression (\ref{msu}) is not that of a
two-loops correction, as it would be for an ${\cal O} \left( S^{-1} \right)$
term, but rather of a truly non-perturbative term from the heterotic
point of view.}.  
Although not exact, expression (\ref{msu}) already contains enough 
information for our purposes,
namely that, besides a dependence on the field $S$, there is also
a dependence on the field $U$, the modulus associated to the complex
structure of $T^2$. This also belongs to the  vector multiplets, and,
under duality, is mapped into a \emph{perturbative} modulus of the heterotic
string. Therefore, the shift doesn't have an action only on the
heterotic dilaton, but also on a perturbative modulus.
This is the reason why the heterotic dual is realized 
by performing a perturbative, explicit operation,
in practice by coupling a twist in the currents with a translation 
in the compact space. From the heterotic point of view,
only the $U$-dependent part of the mass formula is visible, but
this is enough to give a perturbatively non-vanishing mass to the states
of the currents, that therefore do not appear in the massless spectrum.
 
On the type~I side, the situation is different: of the six moduli
in the vector manifold corresponding to the six vector
multiplets originating from the compact space, only three,
those associated to the complex structures of three tori,  are visible. The
other three, namely those associated to the  K\"{a}hler classes,
are projected out by the orientifold operation, that keeps only the momenta
of the lattice of the compact space. If the type~IIA field $U$ is dual
to one of these, the mass expression (\ref{msu}) looks completely
non-perturbative on the type~I side, and we expect to not find
an explicit ${\cal N}=4$ type~I construction without open string massless
sector. On the other hand, if $U$ is dual to a perturbative modulus
of the type~I side, we expect, as in the heterotic string,
to observe that, by construction, the gauge sector is massive.
A type~I model satisfying this requirement can be constructed by 
projecting the type~IIB string by $\Omega^{\cal F}
\equiv (-1)^m \times \Omega$, where $\Omega$ is 
the usual orientifold projection, acting as a world-sheet
parity, and $(-1)^m$ is a shift on the momenta of the lattice associated
to some of the compact coordinates.
The partition function of the closed string sector of this model
is given by the sum of the torus and Klein bottle contributions:
\be
{\cal Z}_{\rm closed} \; = \; {1 \over 2} {\cal T} \,  + 
\, {1 \over 2}{\cal K} \, ,
\label{ztk}
\ee 
where the Klein bottle contribution is given by the trace of $\Omega^{\cal F}$:
\be
{\cal K} \, \equiv \, \langle \Omega^{\cal F} \rangle \, 
= \left( V_8 - S_8 \right) \sum_{\vec{m},\vec{n}} 
\langle \vec{m}, \vec{n} | \Omega^{\cal F} 
| \vec{m}, \vec{n} \rangle
= \left( V_8 - S_8 \right) \sum_{\vec{m}} (-1)^m \langle \vec{m}, \vec{0} | 
\vec{m}, \vec{0} \rangle \, ,
\ee
where $\langle \vec{m}, \vec{n} | \vec{m}, \vec{n} \rangle$ denotes
the lattice character corresponding to the compact space \footnote{$Z_{m,n}$
in the notation of Refs. \cite{ads,adds}.}.
We see that the modification of the orientifold projection due to
the shift $(-1)^m$ does not affect the zero-momentum/winding sector.
In particular, the entire ${\cal N}_4=8$ supergravity multiplet of the 
type~IIB string is projected, as usual, onto an ${\cal N}_4=4$ 
supergravity multiplet plus six vector multiplets. 
Here however some of the states which are not invariant under world-sheet
parity are not projected out, but appear in the spectrum, with
a non-vanishing mass. The lightest among these are found at the
first level, $m=1$, and their mass behaves approximately as
${\hbox  {\sl m}} \sim 1 \big/ R$, $R$ being the compactification
radius of the shifted coordinates.

In order to see the open string contribution, and therefore the gauge 
sector, we must analyze the structure of the ultraviolet divergences
of the Klein bottle.
As usual, these are seen as infrared divergences in the so called 
``transverse channel'', obtained by exchanging the role of world-sheet 
space and time, i.e. by performing a T-duality in the string world-sheet
parameter, $t_{\cal K} \to \ell \equiv 1 \big/ t_{\cal K}$.
By a Poisson resummation, we obtain, for the contribution
of the shifted coordinate:
\be
\sum_{m}\, (-1)^m \langle m | m \rangle  ~
\stackrel{\tau \to \ell}{\longrightarrow} ~
\sum_{\tilde{n}} \, \langle 2 \tilde{n} + 1 | 2 \tilde{n} + 1 \rangle \, .
\ee 
We see that, for finite values of the compactification radii,
there is no zero ``winding'' term, and therefore no infrared divergence too.
As a consequence,
there is no need to add an open string sector: the theory is consistent
with the only two sectors of Eq.~(\ref{ztk}).
As in any freely acting orbifold, in the limit of
vanishing radius of the shifted coordinate, we recover an effective
zero-winding contribution (actually, in this limit all the winding numbers 
collapse to zero). In this limit, the action of the projection is not
anymore free, and, like in ordinary orbifolds, where we expect the appearance
of twisted sectors, here we expect the appearance of an open string sector, 
in effect required by non-trivial infrared divergences.
In the opposite limit of infinite radius we expect that
also states odd under $\Omega$ become massless, leading to the
restoration of the original ${\cal N}_4=8$ string, in an appropriate
decompactification limit.

From this example we learn that, depending on
the way the projection acts on the moduli, the type~I dual of the type~IIA
freely acting orbifold can be either a model without or as well with
open string sector. In this second case, the type~I construction
cannot be perturbatively distinguished from the dual of the $SO(32)$ heterotic
string compactified to four dimensions. 
The difference is in the non-perturbative behavior:
while in one case the gauge sector remains massless even at the 
non-perturbative level, in the other the corresponding states  
have a non-perturbative mass, vanishing in the large-$S$ limit.
This ambiguity is related to the fact that the orientifold construction
is a projection essentially blind to any operation performed on the
windings of the lattice of the type~IIB compact space:
a projection $\Omega^{{\cal F} \, \prime} \equiv (-1)^n \Omega$
would in fact produce the same orientifold as $\Omega$ \footnote{The 
decoupling of the winding numbers can otherwise be seen
as the effect of taking a special limit in the moduli coupled 
to these quantum numbers.}.

To complete the discussion, we must say that the mass of the gravitinos
projected out by freely acting orbifold projections
behaves inversely with respect to the mass of the states of the twisted 
sector. Therefore, if the mass of the latter decreases as the field $S$
increases, the mass of the gravitinos of the broken supersymmetries
instead increases, $m^2_{(3/2)} \sim S_2$ for $S_2 \gg 1$.
This is the reason why we can have a type~I dual, namely a weakly coupled
type~I construction, in which the ${\cal N}_4=8$ supersymmetry is explicitly
reduced to ${\cal N}_4=4$($m^2_{(3/2)} \sim S_2 \gg 1$), 
with an apparently massless open string 
sector, ($m_{\rm open}^2 \sim 1 \big/ S_2 \ll 1$). 

\vspace{1.5cm}

\noindent

\section{\bf ${\bf Z_2 \boldsymbol{\times} Z_2}$ orbifolds: ${\bf {\cal N}_4
    \boldsymbol{=} 2}$}
\label{n=2}

We consider now the ${\cal N}_4=2$ orbifolds.
As we will see, some of them correspond, on the type~IIA side, to 
limits in the moduli space of smooth K3 fibrations. In these cases,
the spectrum contains no states non-perturbative from the heterotic 
point of view.
The most interesting are however the constructions
in which part of the massless spectrum is non-perturbative from the
heterotic point of view. 
The "smooth" cases have mostly already been 
discussed in previous works \cite{fhsv,hmFHSV,gkp,gkp2}. Nevertheless,
we quote them here (sections \ref{11,11}, \ref{3,3}), 
both for completeness, and because we will look
at them from a slightly different perspective, by viewing them as
particular cases of a more general picture, that we can now see
in its entirety. In particular, in Refs. \cite{gkp} and \cite{gkp2},
corresponding to sections \ref{11,11} and \ref{3,3},
only the duality between heterotic and type~II string was considered,
while here we include also the type~I string duals. At that time
these were not known, and their construction is presented here 
for the first time.
The analysis of the cases with non-perturbative massless states
(sections \ref{rank48}, \ref{19,19}, \ref{reducedrank})
constitutes on the other hand the most original contribution of this work,
namely an analysis of type~I/heterotic/type~II duality "beyond K3 fibrations". 
As we will see in section \ref{sbreaking},
under certain conditions the ``hidden'' sectors play a special role in the 
breaking of supersymmetry.

Following Ref. \cite{gkr}, the $Z_2 \times Z_2$ orbifolds of the type~II
string can be classified according to the action of the two $Z_2$ projections,
and to the presence of other, ``semi-freely-acting'' projections, 
that can reduce the number of fixed points.
We will present our analysis by first discussing the cases
in which there are no such additional projections: these
will be discussed in section \ref{reducedrank}.

\subsection{\sl  The "rank 48" vacuum}
\label{rank48}

Our analysis starts with a string vacuum corresponding to the simplest 
${\cal  N}_4=2$ type II orbifold construction. Despite its apparent
simplicity, an analysis of this vacuum turns out to be very complicated, and
it constitutes a key step in order to understand how string--string duality
works in non-trivial cases. This theory possesses in fact four apparently different
realizations: it admits  a type I, a heterotic, and  two type II
constructions, these latter distinguished by the sign of the so called "discrete torsion".  
The two type II realizations are therefore related by an exchange 
between vector and hypermultiplets. In the following, we will discuss how,
despite the fact that in no one of the four constructions all 
of these states appear in the perturbative spectrum, their presence can
nevertheless be traced through their contribution to the string corrections to
effective couplings.
In particular, we will consider the threshold corrections to the coupling of a
special "$R^2$" term, defined in Ref. \cite{gkp}. The four constructions have
a non-trivial overlap in the spectrum, but only through the analysis of all of
them we can have a complete picture. Quite interestingly, the independent computations of the
above mentioned term tell us that, in any of the four realizations, the string
"knows" about the presence of the "hidden" states. 

In order to discuss this
theory, we start first with one of the two type II realizations. We consider
then the type I dual, and discuss how one identifies the heterotic dual.
Several aspects are pointed out, which lead to an interpretation of the 
type I/heterotic duality map for ${\cal N}_4=2$ theories slightly different
from the one usually claimed. Then, in section \ref{3,51} we complete the
picture by considering the "mirror" type II construction. 
Since the analysis is forcedly rather lengthy, 
in order to make reader's life easier, we summarize the main results at the end of the section.

\subsubsection{\it The CY$^{51,3}$}
\label{51,3}

The simplest $Z_2 \times Z_2$ orbifold of the type~IIA string is 
the one in which each of the two $Z_2$ operations acts as a twist on four
compact coordinates. In the case ``without discrete torsion'', we
obtain a massless spectrum corresponding to a Calabi--Yau manifold
with Hodge numbers $(h_{1,1},h_{2,1})=(51,3)$ \cite{vwdt}. 
It has three twisted sectors,
each one having sixteen fixed tori, corresponding to sixteen vector
multiplets. They correspond to the gauge bosons of 
$U(1)^{16} \times U(1)^{16} \times U(1)^{16}$ (see Ref. \cite{gkr}). 

Since the type~IIA string, compactified on $T^2 \times T^4 \big/ Z_2$,
is dual to the heterotic/type~I strings compactified on $T^6$,
we expect that the heterotic and type~I duals of this orbifold
are $Z_2$ orbifolds of $T^6$, namely $T^2 \times T^4 \big/ Z_2$.
Since these are not ``freely acting'' projections, we also expect that,
as suggested by the huge amount of $U(1)$'s in the type~II spectrum, 
in the heterotic and type~I string part of these states are non-perturbative. 
A $Z_2$ orbifold of the type~I string produces the four-dimensional
version of the model first constructed in Refs.~\cite{bs,gp}.
This orientifold has two types of D-branes: 
D9 and a D5 branes, the latter generated by the $Z_2$
operation on four coordinates.
The maximal gauge group is $U(16)_9 \times U(16)_5$, with hypermultiplets
in representations and multiplicities such that the one-loop beta
function coefficients vanish \footnote{One $U(1)$ in each of these
two factors is anomalous, so that at the end the stable configuration should
be a subgroup of $SU(16) \times SU(16)$ \cite{6danom}.}. A proposal for 
the heterotic dual was presented in Refs. \cite{6danom,afiq,ap}:
an orbifold with gauge group $U(16)$, that should correspond to a phase
of the type~I string in which the D5-branes sector is massive.
Although the general idea is correct, in the proposed duality identifications
there are subtleties, that we will now discuss. A closer look will
lead us to conclusions slightly different from what previously proposed 
in the literature.

According to \cite{6danom,afiq,ap},
in six dimensions the heterotic and type I
constructions are argued to be dual when on 
the type~I side the D5-branes are separated
and the corresponding gauge group is broken to $U(1)^{16}$. All these $U(1)$ 
are anomalous and, as discussed in Ref. \cite{6danom}, 
owing to some non-perturbative phenomenon,
the corresponding gauge bosons should acquire a mass.  
Moreover, they would possess a non-vanishing beta-function. 
This duality should then be inherited by the four dimensional theory, upon
toroidal compactification, and extended to include a type ~II dual, 
corresponding to a point in which all the D-branes states
are massive. 
If we now look from the point of view of the type~IIA orbifold
construction, expected to be the dual in four
dimensions, we see that it does not contain 
hypermultiplets charged under the $U(1)$'s, and
moreover the gauge beta functions vanish. 
Therefore, we cannot consider the type~IIA construction as corresponding to
the ``D5'' Abelian point in the moduli space of the type~I model.
However, we argue that this construction is the correct
dual of the heterotic and type I constructions. 
In order to understand the apparent discrepancy, we must keep
in mind that, on the type~IIA side, the spectra that appear
at the orbifold point contain only the states that can be explicitly
constructed with vertex operators of the world-sheet conformal theory,
and they must, by construction, respect the factorization of the 
compact target space into the three tori produced by the
$Z_2 \times Z_2$ projection. As a consequence, not only the 
states charged under the gauge group don't appear at the orbifold point
(they are, as is well known, non-perturbative, D-branes states), 
but also the states that are ``multi-charged'', i.e. charged under 
the groups of two different twisted sectors.
Let's see  what this in practice means:

In the type~II orbifold, each twisted sector contains 16 vector multiplets.
It is easy to check that the spectrum of the twisted sector
does \emph{not correspond}
to the simple projection onto the Abelian subset of the
spectrum of the heterotic $U(16)$ model: if we break the gauge group
to $U(1)^{16}$, by switching on Wilson lines, 
we get in fact the sixteen vector multiplets from the breaking of 
$U(16)$, but also 16 hypermultiplets in the ${\bf 16}$ of
$U(1)^{16}$, from the twisted sector. The beta-function of these
$U(1)$'s therefore does not vanish.
The $U(16)_9 \times U(16)_5$ point of the type~I model, where
the gauge beta-functions vanish,
is therefore the candidate to be the dual of the type~II orbifold 
(the second $U(16)$ would then correspond, on the heterotic side,
to a non-perturbative gauge group, of the ``small instantons'' type).
In fact, of the D-brane states,
only the gauge bosons of the Cartan subgroup, $U(1)^{16}_9 \times
U(1)^{16}_5$, can appear on the type~II side;
the hypermultiplets in the ${\bf 120}$ and ${\bf \overline{120}}$
do not appear, because projected out when the gauge group is
reduced to its Abelian subgroup, and the 256 hypermultiplets
in the D9D5 sectors cannot appear either, because they are bi-charged states.
The open string sector of the type~I model therefore correctly accounts 
for two twisted sectors of the type~II orbifold.
There is however still a  mismatch of sixteen $U(1)$ vector multiplets
of the type~II construction,
that should correspond to a sector non-perturbative in both 
the heterotic and type~I constructions: in total, this theory
should therefore possess three gauge sectors \footnote{In the following,
we will also discuss the sixteen hypermultiplets of the
twisted closed sector of the type~I string.}. A look at the
corrections to the effective coupling of the $R^2$ term indeed indicate that
there are three sectors also on the heterotic and type~I sides.
On the type~II side we have in fact (from Eq.~(5.13)
of Ref. \cite{gkr}, valid for the three orbifold planes):
\be
{ 16 \, \pi^2  \over g^2_{\rm II}} = - 6 \log T_2^{(1)} |\eta (T^{(1)})|^4 - 6
\log T_2^{(2)} |\eta (T^{(2)})|^4 - 6 \log T_2^{(3)} |\eta (T^{(3)})|^4 
+ {\rm n.p.} \, ,
\label{nonso}
\ee 
while on the heterotic side we have:
\be
{ 16 \, \pi^2  \over g^2_{\rm Het}} =  16 \pi^2  \, S_2 -
12 \log T_2 |\eta (T)|^4 - 12 \log U_2 |\eta (U)|^4 
+ {\rm n.p.} \, .
\label{nonsohet}
\ee
This correction has been here computed according to the prescription
presented in Ref.~\cite{gkp}: it corresponds to an $R^2$ amplitude
``corrected'' by subtracting a non-diagonal, $F_{\mu \nu}F^{\mu \nu}$ 
contribution, absent on the type~II side, 
but always present on the heterotic side. After this term has been
subtracted, we end up with a simple integration over the lattice 
of momenta and windings of the untwisted two-torus \footnote{The result
follows immediately from the definitions of Ref. \cite{gkp}
and for standard properties of lattice integrals (see for instance
\cite{dkl}).}.  
As explained in Ref. \cite{gkp}, only after this subtraction the two 
computations can be compared. The precise subtraction is not arbitrary, 
and is fixed by holomorphycity requirements. It is therefore
a rather non-trivial fact that, once the precise normalization
of the amplitude and of the moduli $S$, $T$ and $U$  
has been \emph{independently} fixed, in agreement with Ref. \cite{gkp},
the two results (\ref{nonso}) and (\ref{nonsohet}) match for large $T^{(1)}$,
with the identification of $T^{(1)}$ with $\tau_S \equiv 4 \pi S $, and
$T^{(2)}$, $T^{(3)}$ with $ T$ and $ U$ respectively.
Here however we must account for a further subtlety: with respect to the 
$N_V=8$ case of Ref. \cite{gkp}, corresponding to the CY$^{11,11}$
model of section \ref{11,11}, expression (\ref{nonso}) matches with
(\ref{nonsohet}) only up to an overall factor 2. 
This rescaling of the type~II expression of the effective coupling
had to be expected, because now the gauge group
is realized at the level 1, while in the CY$^{11,11}$ case it is at the 
level 2.
As a consequence, the ``strength'' of the dilaton term is, as in the 
${\cal N}_4=4$ case, one half of that of the CY$^{11,11}$ model.
Effectively things work as if, on the type~II side,
we did not have the 1\big/2 factor due to the reduction of supersymmetry.
In practice, expression (\ref{nonso}) accounts for half of the correct result.
This fact is not just a mere coincidence, but we postpone the explanation 
after the discussion of section \ref{3,51}.

The analogous of expression \ref{nonsohet}, as computed on the type~I side,
reads \footnote{See for instance Ref. \cite{ap}, by combining the
tree level and the one loop result of Eq.~(4.20).}:
\be
{16 \, \pi^2 \over g^2} =  16 \, \pi^2 \, S_2 +  16 \, \pi^2 \, S^{\prime}_2 
- 12 \log U_2 |\eta (U)|^4 + {\rm n.p.} 
\label{nonsoI}
\ee
In all the expressions (\ref{nonso}), (\ref{nonsohet}) and (\ref{nonsoI}),
we have omitted the infrared, cut-off dependent running
and we allowed for the presence of non-perturbative terms.
Expression (\ref{nonso}) suggests that 
each of the three twisted sectors has a ``bare'' coupling parameterized by
the volume form modulus, $T^{(i)}$, of its corresponding fixed torus 
\cite{gkr,48}, and it is natural to identify the fields
$T^{(1)}$, $T^{(2)}$ and $T^{(3)}$ of the type~II construction with the fields
$S$, $T$ $U$ of the heterotic and $S$, $S^{\prime}$, $U$ of the type~I 
constructions \footnote{See Ref. \cite{abfpt} for a definition
of these fields on the type~I side.}. 
From the heterotic point of view,  the appearance in pairs of 
non-perturbative states of the small instantons type, with couplings
parameterized respectively by the fields $T$ and $U$, 
is a consequence of the T-duality symmetry, that exchanges the fields 
$T$ and $U$, implying the presence of both the sectors at the same time
\footnote{As is known, T-duality can be non-perturbatively broken
by instantons \cite{apl}, i.e. by terms suppressed as
${\rm e}^{-S}$ in the large-$S$ limit (weak coupling). 
Small instantons, present instead even at the weak heterotic coupling,
cannot break the $T \leftrightarrow U$ symmetry.} \cite{48,poster}.
From a geometric point of view, the type~IIA orbifold corresponds to a
special point of a K3 fibration, in which the fiber degenerates
and new cycles appear. To be more precise, there appear two
further copies $C^{\prime \, i}$, $C^{\prime \prime \, i}$, 
of 16 cycles of the non-degenerate fiber, namely of those
associated to the moduli $C^i$ of (\ref{prep}). 
If, with an abuse of language, we indicate the cycles with the name
of the corresponding scalars in the vector multiplets,
we can express the intersections as:
\ba
S \cdot T \cdot U & \neq  & 0 \, , \nn \\
&& \nn \\
S \cdot C^i \cdot C^i  & \neq  & 0  \, , \nn \\
&& \nn \\
T \cdot C^{\prime \, i} \cdot C^{\prime \, i}  & \neq  & 0   
\, , \nn \\
&& \nn \\
U \cdot C^{\prime \prime \, i}  \cdot C^{\prime \prime \, i} & \neq  & 0   
\; , ~~~~i=1,\ldots,16 \, ,
\ea
while
\ba
S \cdot C^{\prime \, i} \cdot C^{\prime \, j}  & = &
S \cdot C^{\prime \prime \, i} \cdot C^{\prime \prime \, j} \;  = \; 0 
\, \nn \\
&& \nn \\ 
T \cdot C^{i} \cdot C^{j} &  = &
T \cdot C^{\prime \prime \, i} \cdot C^{\prime \prime \, j} \; = \; 0 \, \nn \\
&& \nn \\ 
U \cdot C^{i} \cdot C^{j} & = &
U \cdot C^{\prime \, i} \cdot C^{\prime \, j}  \; = \; 0 \, .
\ea
Therefore, at this orbifold point, the Calabi--Yau space possesses three
K3 fibrations, with bases associated respectively to $S$, $T$ and $U$.
The prepotential of this theory has the following form:
\ba
{\cal F} & \approx & STU - S \, \sum_i C^i C^i 
- T \, \sum_j C^{\prime \, j} C^{\prime \, j}
- U \, \sum_k C^{\prime \prime \, k} C^{\prime \prime \, k} \nn \\
&& ~~~+ \,{\rm F}_{\rm n.p.} ( {\rm e}^{- S}, T, U, C^i,C^{\prime \, j},
C^{\prime \prime \, k}) \, . 
\label{preprime}
\ea
We argue that the type~IIA orbifold corresponds to a theory
with gauge group $U(16) \times U(16) \times G$,
where $G$ has rank 16, and hypermultiplets
in appropriate representations, such as to cancel the gauge beta functions.
A non-vanishing gauge beta-function would in fact imply a ``mixing'' of the
modulus $S$ with the moduli $T$ and $U$. 
The perturbative $T \leftrightarrow U$ symmetry of the heterotic string,
that we expect to be a symmetry of the full theory in the limit
of large $S$, suggests that the third factor, $G$, of the gauge group, is
another $U(16)$. As discussed in Refs. \cite{48,poster},
gauge bosons with coupling parameterized by $U$ must derive
from tensors of the six dimensional theory obtained by decompactifying the 
two-torus. It is however easy to check that the number of tensors, vectors,
and hypermultiplets of this six dimensional theory cannot be chosen
in such a way as to cancel both the six dimensional $\tr R^4$ anomaly and the
four dimensional one-loop gauge beta-functions.
The point is that, even though, from the heterotic point of view, it
seems that we are allowed to decompactify the theory from four to six
dimensions, this operation appearing to be  innocuous owing to the 
``factorization'' of the two-torus, this is not true for the
full string theory underlying the heterotic construction.
As seen from the dual type~IIA point of view, the decompactification of the 
heterotic two-torus is not at all a ``smooth'' operation:
the torus is in fact non-trivially mapped into the fiber of the
type~II compact space, and its decompactification leads to a singular space.
The ``six dimensional'' heterotic theory looks therefore rather like a
``non-compact orbifold'' \cite{poster}. 
In such a kind of situation, we cannot apply
the rules of a ``smooth'' field theory in order to derive the 
constraints on the number of states. In particular,
the equation derived by imposing the vanishing of the six dimensional 
anomaly doesn't apply to this case \footnote{To give a simple example
of situation in which the non-compact orbifold escapes the rules of a 
smooth construction, let's consider the ten-dimensional type~II string.
It is well known that, owing to the absence of a gauge group, the
Green-Schwarz anomaly must be canceled by a doubling of supersymmetry:
${\cal N}_{10}=2$ instead of ${\cal N}_{10}=1$ as in the heterotic or type~I
string. However, we can consider to project out half of
the supercharges by compactifying the string to six dimensions
on $T^4\big/Z_2$. In particular, this can be done in a ``freely acting'' way,
i.e. without generating a gauge group, by going to four dimensions
and coupling the twist to shifts in the  fourth and fifth coordinate.
We can now take the limit of large $T^4$, with an appropriate
decompactification limit of the fourth and fifth coordinate 
in which the shifts are 
not entirely trivialized. We obtain in this way
an  apparently ``ten dimensional'' theory, but with ${\cal N}_{10}=1$
and without gauge group.}.

Also the identification of the heterotic/type~I map
requires to take into account the symmetries of the heterotic and type~I 
construction, and presents certain subtleties, that now we will discuss. 
In order to do this, we must consider the mirror type~II orbifold obtained
by switching on a ``discrete torsion'' in the $Z_2 \times Z_2$ operation
\cite{vwdt}.
This amounts to reversing, in the twisted sectors, the relative GSO projection 
of one $Z_2$ onto the other.

\subsubsection{\it The ``mirror'', CY$^{3,51}$}
\label{3,51}

By compactification of the type~II string on the ``mirror'' space,
we obtain an orbifold with opposite Euler number, i.e. with
sixteen hypermultiplets in each of the three twisted sectors,
and no vectors, apart from those originating from the projection on the
compact space \cite{vwdt,gkr}. 
The latter gives rise, as before, to 3+1 vector- and 4
hyper-multiplets. Although there exists a smooth K3 fibration
with the same Hodge numbers, we nevertheless argue that
it does not correspond to the orbifold under consideration.
The reason is that, here too, the 48 extra states originate
from the three twisted sectors, sixteen per each. 
For what we said in the previous section,
if this model was an orbifold limit of the K3 fibration,
all these states should originate from only one twisted sector, that we 
would identify with the sector whose ``coupling'' is the heterotic dilaton
$S$. This is not possible, because only states
in the Cartan of a group may appear as elementary states,
built on vertex operators, in the twisted sector of an orbifold of the
type~II string. Moreover, if we compute the corrections to the
$R^2$ term, we find the same expression as (\ref{nonso}), i.e.
a sum over three contributions, that we must interpret as 
the contributions of three ``sectors'' of the theory, two of which
non-perturbative from the heterotic point of view.
All this leads us to argue that the operation that on the type~II side
maps a construction into its mirror, thereby exchanging 
the role of vector and hypermultiplets in the twisted sectors, 
from the heterotic point of view amounts to an ``exchange of role'' of vectors
and hypermultiplets in the gauge bundle and in the non-perturbative 
sectors, without anyway affecting the spectrum. 
To be precise: in the $U(16)$ model we have
vector multiplets in the ${\bf 256}$ (the Adjoint), and
hypermultiplets in the $16 \times {\bf 16}$, in the ${\bf 120}$
and in the ${\bf \overline{120}}$ of the gauge group. 
Of these, only the sixteen bosons in the Cartan of $U(16)$
appeared in the type~II orbifold. 
In the mirror, what happens is that only the sixteen 
hypermultiplets in the Cartan of the $SO(16) \times SO(16)$ symmetry group 
built on the ${\bf 120} \oplus {\bf \overline{120}}$ appear.
The other hypermultiplets cannot appear, for the same reason as before, namely
the fact that they are ``bi-charged'' under two different twisted sectors,
and the gauge bosons cannot either, because the Cartan of this
$SO(16) \times SO(16)$ does not intersect the Cartan of $U(16)$.
The spectrum of the theory therefore is the same as before,
and, as before, the gauge beta function vanishes.
However, from the type~II point of view it corresponds to a different choice
of the vertex operators used in order to describe the elementary
states of the twisted sectors.

The fact that the type~II mirror construction possesses three twisted sectors,
with sixteen hypermultiplets in each, indirectly supports our hypothesis that
also the third sector, hidden also for the type~I string,
consists of a $U(16)$ gauge group with hypermultiplets in
the ${\bf 120}$ and ${\bf \overline{120}}$: precisely the states in the
Cartan subalgebras of these representations appear in the third twisted sector
of the type~II orbifold. In order to have a vanishing gauge beta-function,
we need also in this sector some other 256 hypermultiplets transforming in
the ${\bf 16}$ of the gauge group. The natural candidates
are provided by the hypermultiplets in the twisted sector of the
closed string sector of the type~I model
\footnote{Notice that these sixteen hypermultiplets are naturally charged under
the ``third'', hidden sector. They are therefore ``bi-charged'',
and can never appear in the type~II dual orbifold.}.  
Although there are only sixteen of these states, we argue that
this is only an artifact due to the fact that,
in the closed sector, as in a type~II string orbifold,
there are no vertex operators appropriate to
describe states transforming in a non-Abelian group, as it would be
for the 256 hypermultiplets, transforming naturally in the Adjoint
of an underlying $U(16)$: of these, only the sixteen states of the
Cartan subalgebra are visible. On the other hand,
this type~I model is, by construction, symmetric in the D9 and D5 branes
sectors. Compatibility with the heterotic T-duality ($T \leftrightarrow U$
symmetry) requires therefore that the dual of the heterotic ``$S$ sector'' 
is the ``$U$'' sector of the type~I model. 
The natural choice for the heterotic dual seems therefore to require that
the heterotic fields $T$ and $U$ are dual to the type~I fields $S$ and 
$S^{\prime}$, and the heterotic $T \leftrightarrow U$ symmetry
corresponds to the symmetry between D9 and D5 branes, that would both
correspond to non perturbative sectors of the heterotic string.
With this choice, from the heterotic point of view the
spectrum is exactly the one of the $U(16)$ model, with 
hypermultiplets in the ${\bf 120}$ and ${\bf \overline{120}}$, together
with 256 appearing as $16 \times {\bf 16}$:
these latter cannot in fact
correspond to those in the bi-fundamental of the D9- and D5-branes gauge 
group. In order to see this, we consider the correction to the
effective gauge coupling from the point of view of the effective action.
To this scope we consider the renormalization of vertices of the type
$g W_{\mu} \psi \overline{\psi} $, between a gauge boson of the
sector with bare coupling $g$ and matter spinors belonging to the
hypermultiplets charged also under the gauge group with 
gauge bosons $W^{\prime}_{\nu}$ and bare coupling $g^{\prime}$:
this should account for a part of the string result.

\begin{figure}
\centerline{
\epsfxsize=12cm
\epsfbox{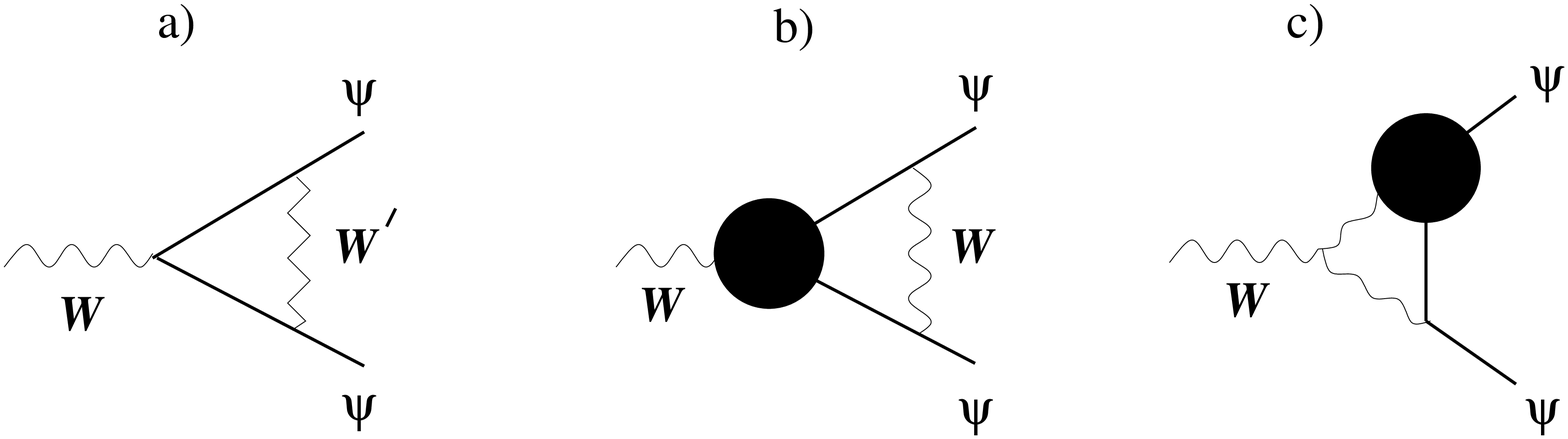}
}
\vspace{0.6cm}

\centerline{
\epsfxsize=11cm
\epsfbox{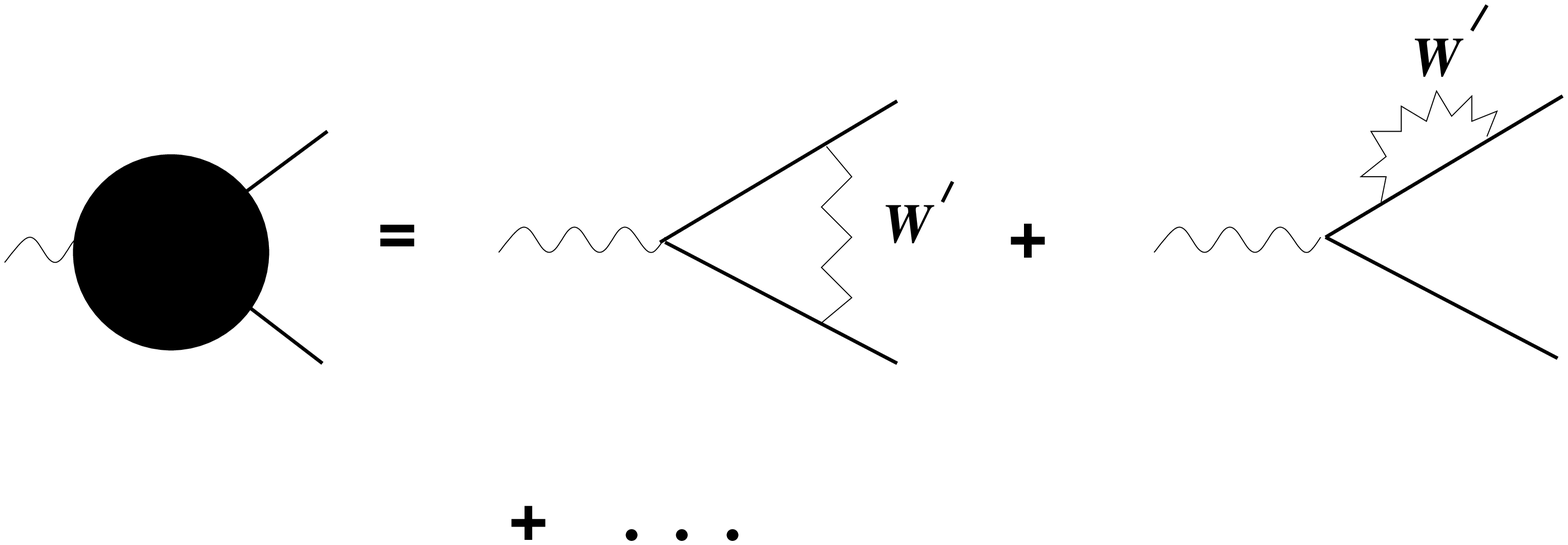}
}
\vspace{0.3cm}
\caption{diagram a) is ``tree level'' from the point of view of the 
coupling $g$, whereas diagrams b) and c) are one-loop.}
\label{wff}
\end{figure}

The contribution of these states to the
renormalization of the coupling of the $U(16)$ gauge group originates,
at the leading orders,
from the diagrams (or better, class of diagrams) of figure~\ref{wff}.
\noindent
All these diagrams give contributions proportional to 
$1 \big/ S^{\prime}_2$. They are therefore forbidden 
by the perturbative symmetries of the ${\cal N}_4=2$ supergravity
coupled to super Yang--Mills. On the other hand, on the type~I side
there is no contradiction, because all these contributions are
non-perturbative from the string point of view. The contradiction would
exist on the heterotic side: in fact, a diagram like a) 
would provide a moduli-dependent, ``tree level'' 
correction to the universal tree level coupling, and diagrams b) and c)
one loop contributions, in a theory with vanishing beta-function.
There are only two possible ways out: either \romannumeral1) the bosons 
$W^{\prime}$ are massive and decouple from the theory; or
\romannumeral2) the hypermultiplets in the ${\bf 16}$ of $U(16)$
are not charged under a ``hidden'' gauge group, i.e. do not coincide
with the 256 hypermultiplets in the D9-D5 sector of the type~I dual.

Solution \romannumeral1) is the one considered in Ref. \cite{ap},
following Ref. \cite{6danom}, where
it is argued that the heterotic theory corresponds to a phase of the 
type~I string in which $U(16)_5$ is broken to $U(1)^{16}$.
All these gauge bosons, since anomalous, should then acquire a mass.
There would therefore be at the end no D5-branes sector.
This hypothesis however hardly accounts for what happens:
its weak point is that, if this was the case,
there would be no D5-branes sector at all. As a consequence, there would
also be no contribution to the 
correction of the $R^2$ term proportional to $S^{\prime}_2$, 
as in Eq.~(\ref{nonsoI}). Therefore, this would not
match with the heterotic counterpart, Eq.~(\ref{nonsohet}).
If on the other hand we suppose that the mass of these bosons is
non-perturbative, behaving as $m \sim {\rm e}^{-S^{\prime}}$ or
$m \sim {\rm e}^{-S}$ (therefore vanishing in the type~I perturbative limit),
we remain with the problem of anomaly cancellation.
\emph{The hypermultiplets of the heterotic twisted sector must therefore
correspond to those in the type~I twisted closed sector},
solution \romannumeral2).
We now come back to the mismatch by a factor 2 between
expression (\ref{nonso}) and (\ref{nonsohet}). As we have noticed, this
is a matter of overall normalization, that can be fixed by adjusting just 
one term. We look therefore at the ``$S$'' term/sector.
The mismatch between heterotic and type~II computations
has its origin in the fact that, on the heterotic side, the model is,
in the above mentioned sense, ``self mirror''. Namely, since it contains
both the vector multiplets of $U(16)$ and the hypermultiplets in the
${\bf 120}$ and ${\bf \overline{120}}$, it accounts for both
the ``$T^{(1)} \leftrightarrow S$'' sectors of the CY$^{51,3}$ orbifold and 
for its mirror, CY$^{3,51}$. Expression (\ref{nonso}),
computed on the  CY$^{51,3}$ model, accounts therefore for just one half of the
complete result.

\vspace{.4cm}

{\bf Summary}

\vspace{.2cm}

\noindent
To summarize: 
this theory has, besides a common ${\cal N}_4=2$ sector, originating from the
projection of the original ${\cal N}_4=8$ supergravity to ${\cal N}_4=2$ and
consisting of the supergravity multiplet plus three vector and four 
hypermultiplets, three gauge sectors, whose ``bare'' couplings are
parameterized by three fields, ``$S$'', ``$T$'' and ``$U$'' respectively,
and gauge group $U(16)_S \times U(16)_T \times U(16)_U$. They appear in the
following way in the various constructions:

\begin{itemize}

\item On the heterotic side only the sector corresponding to $U(16)_S$
is perturbative, with $S$ the dilaton--axion field. 

\item On the type~I side, only $U(16)_T$ and $U(16)_U$, with $T$and $U$
corresponding to the fields parameterizing the coupling of the D9 and D5
branes. 

\item On the type~II side, one can see either
the subgroup $U(1)^{16}_S \times U(1)^{16}_T \times U(1)^{16}_U$,
or, in the mirror orbifold, the 48 hypermultiplets in the ``Cartan''
subalgebras of ${\bf 120}_S \oplus {\bf \overline{120}}_S
 \oplus {\bf 120}_T \oplus {\bf \overline{120}}_T \oplus
{\bf 120}_U \oplus {\bf \overline{120}}_U$.

\end{itemize}

\noindent
Our proposal is that, for this theory, 
duality between type II, heterotic and type I constructions works
already at the orbifold point, once all non-perturbative states are
taken into account; it is not necessary to look
at a point in the moduli space at which
all the gauge states become massive, as previously proposed.
The correspondence of the various sectors and fields is summarized in table
(\ref{table51,3}).
\begin{table}
\begin{center}
\begin{tabular}{c | c | c | c | c | c |}
& {\rm Heterotic} & {\rm Type~I} & {\rm Type~IIA} & {\rm rank} & $R^2$
{\rm behavior }  \\  \hline
{\rm sector/coupling} & $S$ & $U^{\rm I}$ & $T^{(1)}$ &
{\bf 16} & $\eta$ \\ \hline
{\rm sector/coupling} & $T$ & $S^{\rm I}$ & $T^{(2)}$ &
{\bf 16} & $\eta$ \\ 
\hline
{\rm sector/coupling} & $U$ & $S^{\prime \, {\rm I}}$ 
& $T^{(3)}$ &
{\bf 16} & $\eta$ \\ \hline
\end{tabular}
\caption{Duality identifications for the "rank 48" vacuum.}
\label{table51,3}
\end{center}
\end{table}
In this table, we quote only the "twisted sectors" of the theory.
For each sector, we indicate the field parameterizing the gauge coupling
in each of the three string realizations, the rank of the gauge group, and
the behavior of the $R^2$ threshold corrections, schematically
summarized by the relevant Jacobi function appearing in the 
contribution of that sector to the threshold corrections. Therefore, "$\eta$"
stays here for: ${1 / g^2} = \log X_2 |\eta(X)|^4 + \ldots$, where the
field $X$ is specified on the same line for each string realization.
On the heterotic side,
the sectors "${\rm T}$" and "${\rm U}$" are non-perturbative, small
instanton's like. On the type I side, this is the case for 
the sector "${\rm U}^{\rm I}$". The "${\rm S}^{\rm I}$" and   
"${\rm S}^{\prime \, \rm I}$" sectors correspond instead to D9 and D5
branes respectively. As it can be noticed, as opposed to the
${\cal N}_4=4$ case, the heterotic currents don't 
correspond necessarily to the D9 branes sector of the type I dual.

We remark that the type~II vacuum is not
based on a self mirror Calabi--Yau. Therefore, this theory is not protected
against non-perturbative corrections: these could be responsible for certain
mass terms, not appearing at the perturbative level. Among these terms,
we expect those responsible for a mass of the anomalous $U(1)$'s, namely
those whose generator has a non-vanishing trace.

\vspace{.7cm}

\subsection{\sl  The CY$^{19,19}$}
\label{19,19}

\vspace{.4cm}

We move now to a \emph{\bf type~IIA} construction based
on a $Z_2 \times Z_2$ orbifold that can be viewed as a singular limit of
a Calabi--Yau with Hodge numbers (19,19) \footnote{There exists  
a smooth manifold with the same Hodge numbers: the ``Del Pezzo'' surface
(\cite{mvII,cl,acl}). It is not a K3 fibration, and we don't know whether
the orbifold we are considering corresponds to a limit in the moduli space
of this manifold.}.
In this model, of the three effective orbifold projections, 
namely the two $Z_2$ and their product, only one is coupled to a 
translation in the corresponding ''fixed'' tori, and therefore  acts freely 
by lifting the mass of all the associated states.
The other two projections act ordinarily as $Z_2$ twists, 
each one possessing sixteen fixed tori \footnote{For instance,
this orbifold can be constructed with the following operations
on the coordinates of $T^6 = T_{(1)}^2 \times T_{(2)}^2 \times T_{(3)}^2$:
$Z_2^{(1)}$ acts as a twist $x \to - x$ on the last four coordinates:
$x_3, \ldots , x_6$, and as a shift along the momenta of, say, $x_1$, while
$Z_2^{(2)}$ acts as twist $x \to - x$ on the first four coordinates:
$x_1, \ldots , x_4$. Therefore, the product $Z_2^{(3)} = Z_2^{(1)} \times
Z_2^{(2)}$ acts as a twist on $x_1,x_2, x_5, x_6$, and as a shift on
$x_1$. However, as is known, a $Z_2$ shift has a trivial
effect when performed along a twisted coordinate. Therefore,
the action of $Z_2^{(3)}$ is not free.}. Each of the two twisted sectors
provides eight vector multiplets and eight hypermultiplets.
These sectors appear therefore to possess a kind of effective ${\cal N}_4=4$
supersymmetry, at least as far as we look at the multiplicity
of the states.
The correction to the $R^2$ term read \footnote{This can be seen
from Ref. \cite{gkr},
by combining Eq.~(5.12) with the results of lattice integrals, expressions
(E.24), (E.25), for the appropriate orbifold action, quoted in table (D.1).}:
\be
{16 \, \pi^2 \over g^2} = -2 \log T_2^{(1)} | \theta (T^{(1)})|^4 -6
\log T_2^{(2)} | \eta (T^{(2)})|^4 -6 \log T_2^{(3)} | \eta (T^{(3)})|^4 \, ,
\label{tee}
\ee
where for simplicity we don't show the infrared running. 
In this case expression (\ref{tee}) is expected to be exact, because
the compact space is self-mirror.
The first term, with a theta function, accounts for the contribution of
the massive twisted sector, associated to the freely acting projection.
Which one of the three theta functions is really involved
depends on the specific choice of translation inside the torus.
All choices are equivalent up to an $SL(2,Z)$ transformation of the
moduli of the torus
\footnote{Also in the following, whenever a specification of
the exact action of the shift on the torus is not required for the purpose
of the discussion,  we will use the generic
symbol "$\theta$", intended to stay for one of the three Jacobi functions,
$\theta_2$, $\theta_3$, $\theta_4$.} (see for instance Ref. \cite{gkp},
appendix E for a more detailed account).
The second and third term account for the contribution of
the non-freely acted twisted sectors.
As in the previous case, the correction is a sum of three terms,
that depend on fields that we interpret as the ''bare'' coupling moduli of 
the three corresponding sectors. 
We observe therefore  that, whenever, owing to the free action 
of the corresponding orbifold projection, the states of a certain sector
acquire a mass, the contribution of the corresponding coupling modulus
appears through a function theta instead of eta. This is related to the fact
that a shift in the lattice of a torus breaks the initial $SL(2,Z)$
symmetry of the corresponding moduli to a specific $\Gamma_2$
subgroup, the one preserved by the theta function 
\footnote{More details about the relation between the subgroup of 
$SL(2,Z)$ preserved by a given lattice shift and what is the Jacobi
function appearing in the threshold correction, after a summation over the
lattice of shifted momenta/windings, can be found in the Appendix C of
Ref. \cite{6auth}, Eq.~(C.17) and followings.}.
The symmetry between $T^{(2)}$ and $T^{(3)}$ suggests that,
if the three moduli $T^{(i)}$ have in some way 
to correspond to the three moduli
$S$, $T$, $U$ of a heterotic dual, we have to expect
that the map is between the second and third 
and the heterotic moduli $T$ and $U$, while $T^{(1)}$ should correspond to the
heterotic dilaton $S$. On the heterotic side, in fact, an orbifold projection
cannot break the $SL(2,Z)$ invariance of $T$ and not of $U$ or vice-versa.   
However, a look at the (approximate) mass formula for the missing states of 
this freely acting orbifold \footnote{We recall that this is
obtained by inserting in the ${\cal N}=8$ BPS mass formula the new values of 
momenta and winding numbers, as derived from the orbifold action.} gives:
\be
m^2 \sim {1 \over T_2^{(1)} U_2^{(1)}} \, .
\ee
Translated into heterotic fields, this would read:
\be
m^2 \sim {1 \over S_2 Y_2^H} \, ,
\label{msy}
\ee
where $S$ is the dilaton--axion field and $Y^H$ a modulus in the
hypermultiplets. But, on the heterotic side, the first modulus
is non-perturbative, and the second twisted.
Therefore, the type~IIA super-Higgs mechanism underlying this construction
would not be visible.  
This means that, apparently, we would have massless gauge bosons
from the currents. In reality, they would have a non-perturbative mass,
vanishing in the limit of large-$S$.

An indication that this may be the case is provided by
the analysis of the \emph{\bf type~I} string. Indeed, for large $T^{(1)}$
and $T^{(2)}$, (\ref{tee}) behaves like:
\be
{16 \, \pi^2 \over g^2} \sim 2 \, \pi \, T_2^{(2)} \, + \, 2 \, \pi \,
T_2^{(3)} \, - 2 \log T_2^{(1)} | \theta (T^{(1)})|^4 \, .
\ee
This is exactly the correction we would compute in the so called
''winding breaking'' model of Ref. \cite{adds},
with the identifications $T^{(2)} \leftrightarrow 2 \tau^{\rm I}_S$,
$T^{(3)}  \leftrightarrow 2 \tau_{S^{\prime}}$ and 
$T^{(1)} \leftrightarrow U$, where as usual 
$\tau^{\rm I}_S \equiv 4 \pi S^{\rm I}$, 
$\tau_{S^{\prime}} \equiv 4 \pi S^{\prime}$ are the complex moduli
associated to the couplings of the D9 and D5 branes respectively. 
The ``winding breaking'' construction is
a freely acting $Z_2$ orbifold of the ${\cal N}=4$ type~I string, in which
the translation associated to the orbifold twist acts on the
windings numbers of the fixed torus, something that 
is not expected to be visible on the heterotic side.
The symmetry between the two massless twisted sectors of the type~IIA
orbifold would therefore correspond to the symmetry between
D9 and D5-branes sectors. And indeed, in these sectors the rank
of the gauge group is half of the ordinary one, with an equal number
of vector and hypermultiplets (see Ref. \cite{adds}). This means that,
at the Abelian point, the type~I construction has exactly the same massless
spectrum of the type~IIA orbifold based on this CY$^{19,19}$.
The factor 2 in the normalization of the map between the
moduli $T^{(2)}$, $T^{(3)}$ and the couplings of the D9- and D5-branes sectors
precisely accounts for the rank reduction/level doubling 
of the gauge algebra in these sectors. An analogous factor will appear
in the CY$^{11,11}$ construction of the next section \footnote{For 
a discussion, see also Ref. \cite{hmFHSV}.}.

\begin{figure}
\centerline{
\epsfxsize=11cm
\epsfbox{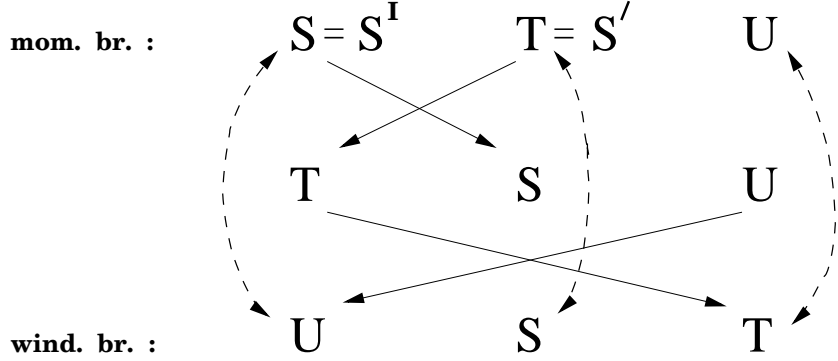}
}
\vspace{0.4cm}
\caption{In the first line, we have 
identified heterotic and type~I fields as for the
momentum breaking duality: $S^{(\rm het)} \equiv S$ with the type~I 
field $S^{\rm I}$, $T^{(\rm het)} \equiv T$ with $S^{\prime}$, while
$U^{(\rm het)} \equiv U$ is mapped into the $U$ field of the type~I model.
Dashed lines indicate the final duality map 
from the heterotic and type~I parameters.}
\label{stu}
\end{figure}

In order to see the relation of this construction with the {\bf heterotic} 
string, we go back to the ${\cal N}=4$ level, the point at which we know that 
the two theories are dual. At this stage, a simple reduction of supersymmetry
through a freely acting orbifold projection, lifting the mass of 
certain states without adding twisted sectors, should adiabatically preserve
the duality between the two constructions. This is in fact the case, as we will
see, of the ``momentum breaking'' construction, in which a freely acting 
$Z_2$ projection acts as a twist on four coordinates, and as a translation
on the two torus, resulting in a shift of the momenta (see Ref. \cite{adds}). 
The latter just label the quantum numbers of the Kaluza--Klein states, indeed
present in any simple dimensional reduction of the 
effective supergravity theory underlying string theory. 
Duality with the heterotic string then works just because it is a
supergravity duality, that does not involve pure stringy states.    
This is the starting point, or in other words the situation 
one has to take as a reference point in order to understand the duality
between ${\cal N}=2$ heterotic and type~I constructions. 
In order to understand how things work in the case of ``winding'' breaking,
one has to reduce the model to a ``momentum breaking'' situation, 
namely to a case in which the orbifold acts on the Kaluza--Klein states. 
Therefore, first of all one has to convert the windings into momenta.
This is consistently achieved after a T-dualization  of coordinates,
that involves the exchange of the fields $S^{\rm I}$ and $S^{\prime}$,
i.e. the D9 and D5-branes sectors \footnote{From the type~I point of 
view the theory is symmetric in these two sectors,
and performing or not this exchange is irrelevant. But 
only with this exchange we get a consistent map, that
respects T-duality of the heterotic string.}, and an exchange of  
K\"{a}hler class and complex structure in the fixed torus.
In other words, we use the fact that a winding shift on the type~IIB
(and type~I) string is equivalent to a momentum shift on the type~IIA
(and type~I$^{\prime}$) string. 
At the end of the day, this operation has mapped the field $S^{\rm I}$ into 
the field $U$, the field $S^{\prime}$ into the field $S^{\rm I}$ and the field
$U$ into the field $S^{\prime}$, as can be checked from figure~\ref{stu}.
%\vspace{0.5cm}
%\vspace{0.6cm}
%\noindent
This means that now, as compared to the ordinary duality, it is 
the field $U$ which has the role of the heterotic dilaton,
the field $S^{\rm I}$ which is dual to the heterotic field $T$
\footnote{In the ``momentum breaking'' model, the relation is the
``ordinary'' one, that identifies the states of the currents of the 
heterotic string with those of the D9-branes of the type~I string,
and therefore maps the heterotic dilaton $S^{(\rm het)}$ into
the type~I field $S^{\rm I}$, the volume modulus of the heterotic
two-torus, $T^{(\rm het)}$, into the type~I field $S^{\prime}$,
and the complex structure modulus $U$ into the complex structure
modulus of the type~I side.}, and the field $S^{\prime}$ that is dual
to the heterotic field $U$. Correspondingly, now the sector of D9-branes,
whose coupling is parameterized by the field $S^{\rm I}$ of type~I, corresponds
to the heterotic small instantons, with coupling $T$, and the sector
of D5-branes corresponds to the vectors corresponding to the 
non-perturbative tensors of six dimensions. 
This construction is then an example of heterotic \emph{perturbative}
${\cal N}_4=2$ theory in which the entire gauge group is
non-perturbative.
The mass of the states in the perturbative sector, i.e. the 
states originating from the currents, vanishes in the large-$S$ limit.
This however does not mean that these states are necessarily
extremely light in the weak heterotic coupling: 
as is clear from Eq.~(\ref{msy}),
besides a dependence on $S$, their mass carries also a dependence on
another, perturbative field, $Y^H$, that can weaken the suppression
due to the large value of the dilaton.
On the type~IIA side, the geometric space corresponding to this 
compactification is a ``singular'' K3 fibration similar to that discussed in 
Section \ref{51,3}, i.e. possessing other two K3 fibrations, that appear
at points in which the fiber degenerates.
In this case, however, the cycles associated to $C^i$, $i=1,\ldots,16$,
dual to the heterotic perturbative vector multiplets of the currents,
are absent.

Although on the heterotic side the operation that lifts the mass of the states 
originating from the currents looks entirely non-perturbative,
so that we cannot identify the heterotic dual from
its perturbative gauge spectrum,
we nevertheless have some further indication that allows us to 
discriminate what is the explicit heterotic construction this theory  
corresponds to.
The left-right symmetric, non-freely acting $Z_2$ orbifolds of the 
$E_8 \times E_8$ heterotic string can be divided in two classes, 
distinguished by the type of embedding of the spin connection into 
the gauge group: the 24 instantons of the K3 can in fact be
embedded either symmetrically or asymmetrically in the two $E_8$ factors.
The model described in Section \ref{51,3} is in the moduli space of the 
constructions with (12,12) embedding described in Refs.~\cite{6danom,afiq},
that belong to the first class, 
while to the second belong compactifications that
lead to $SU(2) \times E_7 \times E_8$ or similar gauge groups.
In Ref. \cite{gmonobis} we noticed that the symmetrization
in the two $E_8$ factors, obtained by starting from an ordinary,
asymmetric embedding, and adding a special Wilson line (see also 
\cite{6danom,afiq}), ``doubles'' the
number of hypermultiplets originating from the twisted sector
\footnote{In the sense that, at the Abelian point, the number of
charged hypermultiplets in the symmetric embeddings is twice
the one in the asymmetric embeddings.}.
This suggests that the heterotic dual of the CY$^{19,19}$,
in which each twisted sector corresponds to a ``level 2''
realization, $N_V=N_H=8$, should be sought for
in the orbifolds with asymmetric embedding \footnote{ There actually exist  
also other orbifolds in which the number of twisted hypermultiplets is reduced
by half, with respect to the (12,12) orbifold. 
This is however obtained as the effect of a freely
acting, rank-reducing projection, and they have  threshold corrections that
don't match with those computed on the type~II side.
They  must therefore be excluded: they indeed correspond to other 
constructions, and will be discussed in Section \ref{reducedrank}.}.  
This hypothesis is supported by the analysis of the corrections
to the ``modified $R^2$ term'', defined in Ref. \cite{gkp}.
In the case of symmetric embedding, they are given by (\ref{nonsohet}).
In order to obtain their expression for the case of asymmetric embedding, 
we have just to keep in mind that, owing to uniqueness properties
of modular functions, once \emph{summed over the orbifold
boundary conditions}, the integrand of the partition function with
the insertion of the appropriate operator corresponding to the amplitude
under consideration must be proportional to that of the models with
symmetric embedding \cite{kkpr}. In particular, 
owing to the properties of the $R^2$ amplitude, that amounts
essentially to the insertion of the modular function $E_2$ in
the partition function, this turns out to be valid
separately for the various terms composing this amplitude,
corresponding to the insertion of the operators ``$R^2$'', 
``$F_{\mu \nu}F^{\mu \nu}_{(\rm torus)}$'' and 
``$F_{\mu \nu}F^{\mu \nu}_{(\rm gauge)}$'' \footnote{We
refer to Ref. \cite{gkp} for a detailed definition of these operators.}.  
In order to fix the normalization, we just consider that the final result
does not depend on the details of the ``internal'' sector
with central charge $c=(4,20)$, corresponding to the four twisted
bosons and to the currents. In particular, the result is the same as
at the $U(1)^{16}$ point. There, the coefficient of the $q^0$ term of
``$F^2_{\rm gauge}$'', proportional to the gauge beta function, is 
given, up to a universal normalization, just by the number of twisted 
hypermultiplets, charged under the
$U(1)$'s. We know that these are half of those of the symmetric embeddings.
Therefore, the overall normalization of the amplitude is one half
of that of the symmetric embeddings.
As a consequence, the effective coupling of the ``modified $R^2$ term'' reads:
\be
{16 \, \pi^2 \over g^2} = 16 \, \pi^2 \, S_2 \, 
- 6 \log T_2 | \eta ( T) |^4 \, - 6 \log U_2 | \eta ( U) |^4 \, + {\rm n.p.}
\, .
\label{nonsoasym}
\ee
The second and third term on the r.h.s. account for the one-loop contribution,
and match with the second and third term of (\ref{tee}), with the
identifications: $T \leftrightarrow T^{(2)}$, $U \leftrightarrow T^{(3)}$. 
The tree level contribution, linear in the dilaton field, $S_2$,
should then correspond to the large $T^{(1)}$ limit of the first term in
(\ref{tee}). According to this string--string duality  scenario, 
in the heterotic constructions with an embedding of the spin connection
asymmetric in the two $E_8$,
the sector of the perturbative currents of the heterotic 
string is non-perturbatively unstable, and the corresponding states are
all lifted to a non-zero mass. This non-perturbative instability
has a counterpart in the problems that arose in Ref. \cite{dmw}, when
trying to understand the behavior of this sector for finite values of the 
dilaton.

As we did at the end of section \ref{rank48}, we summarize the
main duality identifications and properties of this theory in a table:
\begin{table}[here]
\begin{center}
\begin{tabular}{c | c | c | c | c | c |}
& {\rm Heterotic} & {\rm Type~I} & {\rm Type~IIA} & {\rm rank} & $R^2$
{\rm behavior } \\ \hline
{\rm sector/coupling} & $S$ & $U^{\rm I}$ & $T^{(1)}$ &
{\bf 0} & $\theta$ \\ \hline
{\rm sector/coupling} & $T$ & $S^{\rm I}$ & $T^{(2)}$ &
{\bf 8}$\vert_2$ & $\eta$ \\ \hline
{\rm sector/coupling} & $U$ & $S^{\prime \, {\rm I}}$ 
& $T^{(3)}$ & {\bf 8}$\vert_2$ & $\eta$ \\ \hline
\end{tabular}
\end{center}
\label{table19,19}
\caption{The duality identifications of the "$19,19$" vacuum.}
\end{table}

\noindent
With respect to Table  (\ref{table51,3}), here we introduced a further
notation: the subscript "$|_2$" indicates that the gauge group is 
realized at the level 2, with an equal number of vector and hyper multiplets
in the same representation. 
As we remarked, from the heterotic point of view matter and gauge
states are here entirely non-perturbative.

\subsection{\sl The CY$^{11,11}$}
\label{11,11}

This type~IIA construction and its duality with the heterotic string
has been widely investigated (see Refs. \cite{fhsv,hmFHSV,gkp}).
On the type~IIA side, the $Z_2 \times Z_2$ orbifold point
is constructed by coupling one of the two orbifold projections to a translation
along two tori, so that only one twisted sector possesses fixed points,
and provides massless states; the other two are massive.
Supersymmetry is spontaneously broken from ${\cal N}_4=4$
to ${\cal N}_4=2$. The two gravitinos corresponding to the spontaneously
broken supersymmetries have masses behaving approximately
as \cite{gkp}:
\be
m^2_{(3/2)} \sim {1 \over T_2^{(2)} U_2^{(2)}} 
+{1 \over T_2^{(3)} U_2^{(3)}} \, .
\label{label}
\ee 
The heterotic dual is constructed as a $Z_2$ freely acting orbifold
coupled to a twist in the currents, that
produces a level-two realization of the algebra: the result is 
a gauge group of rank 8, and an equal number of vector and hypermultiplets
(for details, see Refs. \cite{fhsv,gkp}).
Under string--string duality, the type~IIA fields $T^{(2)}$ and 
$T^{(3)}$ are mapped respectively to
the K\"{a}hler class and complex structure moduli, $T$ and $U$, 
of the heterotic untwisted torus. 
From the heterotic point of view, the mass of the
gravitinos corresponding to the broken supersymmetries
appears as (see Ref. \cite{gkp}):
\be
m^2_{(3/2)} \sim {1 \over T_2 U_2} \, ,
\ee
(notice that the contribution of the moduli in the hypermultiplets
does not appear, being these fields twisted on the heterotic side).
The full correction to the $R^2$ term, at the orbifold point, reads 
(Ref. \cite{gkp}, Eq.~(2.18)):
\be
{16 \, \pi^2 \over g^2} = -6 \log S_2 |\eta (S)|^4 -2 
\log T_2 |\theta (T)|^4 -2  \log U_2 |\theta (U)|^4 \, , 
\label{r88}
\ee
where, as usual, we omit the infrared running. 
The normalization of the second and third term allows to identify the 
leading contribution of the moduli 
$T$ and $U$ with the ``gravitational beta function'' of field theory.

In order to find the type~I dual, once again we consider the 
expressions for the mass of the gravitinos. Although approximate 
(they should actually be given in terms
of modular functions of the fields, respecting the duality group
preserved by the orbifold projection), they already tell us that
there are essentially two possibilities: either one of the
two moduli in the vector multiplets entering in the type~II gravitino
mass expressions corresponds to the only perturbative modulus
of the vector multiplets of the type~I string, or both they
correspond to a non-perturbative, ``coupling constant'' modulus. 
There are no other possibilities,
because in this theory there are only three moduli in
the vector multiplets \footnote{The Wilson lines are here
fixed to discrete values.}, and two of them are the fields 
$S^{\rm I}$ and $S^{\prime}$. 

In the first case, the mechanism of spontaneous breaking of supersymmetry
should be visible also on the type~I side: 
in this case, the type~I dual would appear as a ``winding breaking'', 
in which one of the two D-branes sectors, although
perturbatively massless, is in reality massive. However, the only
``winding breaking'' type~I construction is the one of section \ref{19,19},
which has $N_V=N_H$ and all the characteristics to be the ``finite'' theory
we argued. 

In the second case, the type~I dual
is constructed by applying the ``winding breaking'',
or ``M-theory breaking'' freely acting $Z^{\cal F}_2$ 
projection of Ref. \cite{adds},
to the $\Omega^{\cal F}$, freely acting orientifold of Section \ref{sbroken}.
If the shift of $\Omega^{\cal F}$ acts along one of the coordinates
twisted by $Z^{\cal F}_2$, it is not visible in the 
$Z^{\cal F}_2$-twisted/projected sector. This means that the D5-branes 
sector is the same as in the ``winding breaking'' model of 
Section~\ref{19,19}. On the other hand, for the
reason explained in Section \ref{sbroken}, the D9-branes sector is missing.
The D-branes spectrum possesses therefore an effective 
${\cal N}_4=4$ supersymmetry,
with a gauge group realized at the level 2, with rank 8.
There is on the other hand no deep reason why the perturbative sector of the 
heterotic string
appears on the D5-branes of the type~I dual: indeed, in the $Z_2$ orbifolds
of the type~I string, the D9 and D5 branes are essentially on the same footing.
In order to obtain the gauge sector on the D9-branes,
one just has to project the type~IIB string by 
$\tilde{\Omega}^{\cal F} \equiv \Omega^{\cal F} \times I_4$, where
$I_4$ is the reflection, $x \to -x$, along the four coordinates
twisted by $Z_2^{\cal F}$.
For a quick overview, we summarize the results in tables~(\ref{table11}),
representing the two equivalent possibilities.
\begin{table}[here]
\begin{center}
\begin{tabular}{c | c | c | c | c | c |}
& {\rm Het.} & {\rm Type~I} & {\rm Type~IIA} & {\rm rank} & $R^2$
{\rm behavior }  \\ \hline
{\rm sector/coupling} & $S$ & $S^{\rm I}$ & $T^{(1)}$ &
{\bf 8}$\vert_2$  & $\eta$ \\ \hline
{\rm sector/coupling} & $T$ & $S^{\prime \, {\rm I}}$ 
& $T^{(2)}$ &
{\bf 0} & $\theta$ \\ \hline
{\rm sector/coupling} & $U$ & $U^{\rm I}$ 
& $T^{(3)}$ & {\bf 0} & $\theta$ \\ \hline
\end{tabular}
\end{center}
\vspace{.3cm}
\begin{center}
\begin{tabular}{c | c | c | c | c | c |}
& {\rm Het.} & {\rm Type~I} & {\rm Type~IIA} & {\rm rank} & $R^2$
{\rm behavior }  \\ \hline
{\rm sector/coupling} & $S$ & $S^{\prime \, {\rm I}}$ 
& $T^{(1)}$ & {\bf 8}$\vert_2$  & $\eta$ \\ \hline
{\rm sector/coupling} & $T$ & $S^{\rm I}$ 
& $T^{(2)}$ &
{\bf 0} & $\theta$ \\ \hline
{\rm sector/coupling} & $U$ & $U^{\rm I}$ 
& $T^{(3)}$ & {\bf 0} & $\theta$ \\ \hline
\end{tabular}
\end{center}
\label{table11}
\caption{The two equivalent duality identifications for the "$(11,11)$"
vacuum. The two tables differ for the exchange of $S^{\rm I}$ 
and $S^{\prime \, {\rm I}}$.}
\end{table}
Notice that, as opposed to the non-freely acting orbifolds of sections
\ref{rank48} and \ref{19,19}, here the heterotic currents do
correspond to the D9 (or D5) branes of the type I dual, as they do for
${\cal N}_4=4$.
This is due to the fact that here the supersymmetry breaking projection
acts freely. Therefore, the ${\cal N}_4=2$ theory is a true
projection of the ${\cal N}_4=4$ one. There are no additional (massless)
sectors, and the ${\cal N}_4=2$ string string duality map is directly
inherited from ${\cal N}_4=4$.

\vspace{.7cm}

\subsection{\bf {\sl The CY$^{3,3}$}}
\label{3,3}

\vspace{.4cm}

Our next step is to go to a more complicated orbifold,
namely the one obtained on the type~IIA side by coupling each $Z_2$ twist to
a $Z_2$ translation in the corresponding fixed torus.
As a result, there are no more fixed tori, and all the three twisted 
sectors are massive (see Refs. \cite{sv,gkp2} 
for details about this construction).
Although here the projection is more involved, this model, 
as the one of section \ref{51,3}, 
possesses a symmetry in the three twisted sectors, that makes
irrelevant the identification of which sector is really going into
perturbative or non-perturbative sectors of the 
dual heterotic and type~I strings.
The correction to the coupling of the $R^2$ term reads 
(Ref. \cite{gkp2}, Eq.~(2.6)):
\be
{16 \, \pi^2 \over g^2} = -2\log T_2^{(1)} |\theta (T^{(1)})|^4
-2 \log T_2^{(2)} |\theta (T^{(2)})|^4 
-2 \log T_2^{(3)} |\theta (T^{(3)})|^4 \, ,
\label{000}
\ee
As always, we omitted the cut-off dependent infrared running.
Unlike in \ref{rank48}, and as in \ref{19,19} and \ref{11,11}, expression
(\ref{000}) is supposed to be exact, because the manifold corresponding
to this orbifold, CY$^{3,3}$, is self-mirror. We expect therefore
both the vector and hypermultiplets moduli spaces to be exact.
Here too a look at the (approximate) mass formula for the 
states of the twisted sectors, whose mass has been now lifted, 
reveals that a super-Higgs
phenomenon is at work (as always in freely-acting orbifolds).
For any of such states, the mass is given as a function
of two moduli. A typical situation is:
\ba
m^2_{(1)} & \sim &  T_2^{(1)} U_2^{(1)}  \, , \nonumber \\ 
m^2_{(2)} & \sim &  T_2^{(2)} U_2^{(2)}  \, , \label{m123} \\ 
m^2_{(3)} & \sim &  T_2^{(3)} U_2^{(3)}  \, , \nonumber  
\ea 
where the indices (1), (2), (3) refer to the three twisted sectors,
and their corresponding tori. The moduli $T^{(i)}$ are associated to
the K\"{a}hler classes, while the moduli $U^{(i)}$ to the complex structures.
For the gravitinos corresponding to the broken supersymmetries,
we have instead (see Ref. \cite{gkp2}):
\ba
m_{(1)}^2 & \sim & {1 \over T^{(1)}_2 U^{(1)}_2} +{1 \over T^{(2)}_2 U^{(2)}_2}
\, ; \nonumber \\
m_{(2)}^2 & \sim & {1 \over T^{(1)}_2 U^{(1)}_2} +{1 \over T^{(3)}_2 U^{(3)}_2}
\, ; \label{m2t2u} \\
m_{(3)}^2 & \sim & {1 \over T^{(2)}_2 U^{(2)}_2} +{1 \over T^{(3)}_2 U^{(3)}_2}
\, , \nonumber 
\ea
The $T^{(i)}$ are moduli in the vector multiplets;
therefore, they are mapped to the heterotic moduli $S$, $T$, $U$,
and to the type~I moduli $S^{\rm I}$, $S^{\prime}$ and $U$.
The $U^{(i)}$ are instead moduli in the
hypermultiplets: they cannot be observed on the heterotic and type~I
sides, being either twisted or projected out by the orientifold action.

\subsubsection{\it the Heterotic dual}
\label{htd}

We want to see how the above construction appears on the heterotic side.
From (\ref{m123}) we see that no one of the three twisted sectors, that
in Section (\ref{51,3}) we have put in relation respectively with the  
perturbative and two non-perturbative gauge sectors of the heterotic string,
is now massless. In each sector the mass of the
states depends on the corresponding coupling constant, and on a
modulus in the hypermultiplets. On a heterotic
orbifold the latter are however twisted. 
Therefore,  on the heterotic side the operation that lifts the mass
of the states on the perturbative sector, the currents,
is not explicitly observable in the mass formulas. 
It is in fact not observable in the ``$S$'' sector, because the mass 
of the states look as entirely non-perturbative;
it is not ``observable'' in the ``$T$'' and ``$U$'' sectors  
either, because all the states in these sectors are non-perturbative
for the heterotic string. The effect of this operation can
only be indirectly traced in the behavior of certain threshold corrections,
not at the level of the perturbative spectrum. On the other hand, from
(\ref{m2t2u}) we see that anyone of the gravitino mass formulas 
depends on at least one modulus which is perturbative on the heterotic side.
We expect therefore the heterotic dual of this construction
to be a freely acting orbifold in which, apparently, supersymmetry
is spontaneously broken from ${\cal N}_4=4$ to ${\cal N}_4=2$, 
but in which actually also the breaking from ${\cal N}_4=8$
to ${\cal N}_4 = 4$ is spontaneous. 
We expect also to find a perturbatively massless 
gauge sector, originating from the currents.

All the geometric freely acting orbifolds considered 
in Ref. \cite{kkprnew} fulfill this
requirement. We argue that, at least in  the models that
correspond to a compactification of the heterotic string on an elliptically
fibered K3, there exist no massless states from the currents.
This does not mean that all the cases included in the analysis
of Ref. \cite{kkprnew} belong to this class: they may well
correspond to heterotic vacua in which it is not possible to
identify such a structure, and therefore may not correspond
to a perturbative type~II vacuum.

It exists however also a heterotic orbifold in which
the mass of the states from the currents is explicitly, perturbatively
lifted to a non-zero value \cite{gkp2}.
And, as correctly argued in Ref.~\cite{gkp2}, indeed this
model possesses a type~IIA dual, that appears precisely as the orbifold
we have just considered. How is it possible?

In order to understand what is going on, let's look at the heterotic 
freely acting orbifold of Ref. \cite{gkp2} more in detail.
The perturbative operation that twists all the states in the currents
reflects into a dependence of the masses of such states on
both the moduli $T$ and $U$. This means that, if we consider
that the heterotic\big/type~IIA map is $S \to T^{(1)}$, $T \to T^{(2)}$, 
$U \to T^{(3)}$, the mass formula in the first line of (\ref{m123}),
corresponding to this sector, turns out
to depend also on the moduli $T^{(2)}$, $T^{(3)}$.
But these are twisted in this sector, and therefore a dependence on them
can never be observed on the type~II side! 
This moduli dependence appears when the $Z_2$ projection that twists
the planes (2) and (3) acts as a shift not only in the plane (1)
but also in the planes (2) and (3). Despite this further shift,
from the type~IIA point of view the mass formula remains the same. 
Although not directly observable from the mass formulae,
such a ``hidden'' action has a simple explanation, and can be traced
in the construction of this type~IIA orbifold.
It is  therefore worth to go back to some aspects of the construction, 
presented in full detail in Ref. \cite{gkp2}.
There are two freely acting $Z_2$ projections, $Z_2^{(1)}$ and $Z_2^{(2)}$.
They naturally divide the compact space $T^6$ into three tori, or planes:
$T^6 = T^2_{(1)} \times T^2_{(2)} \times T^2_{(3)}$.
$Z_2^{(1)}$ twists the planes $T^2_{(2)}$ and $T^2_{(3)}$,
while $Z_2^{(2)}$ twists the planes $T^2_{(1)}$ and $T^2_{(3)}$.
Therefore, their product $Z_2^{(3)}  \equiv Z_2^{(1)} \times Z_2^{(2)}$
twists the planes $T^2_{(1)}$ and $T^2_{(2)}$.
$Z_2^{(1)}$ and $Z_2^{(2)}$ act necessarily as a shift on the plane they 
leave untwisted.
However, in order to ensure that also $Z_2^{(3)}$ acts freely
(otherwise we would have a massless twisted sector), one of them
must act as a shift also on the third plane. 
Consistency with duality to a perturbative 
heterotic construction requires to pair the translation
on the third plane to an analogous action
on the second one (in order to respect the perturbative
T-duality of the heterotic string,
that reflects in the exchange of $T$ and $U$) \footnote{Notice that
we are not requiring the shift to be identical in both the planes.}.  
Now, depending on whether the shift on the third plane is 
produced by $Z_2^{(1)}$ or by $Z_2^{(2)}$, we have a different
heterotic dual. In the first case, we get the type~IIA dual
of the heterotic orbifold with a perturbative twist of the currents.
In the second case, it is not the first line of (\ref{m123})
that contains a ``hidden'' modification, but the second one 
\footnote{Notice also that necessarily the third formula always contains 
a ``hidden'' modification of this kind.}. 
From the heterotic point of view, the dual
of this second situation is therefore a freely acting orbifold in which
the states of the currents are apparently massless.

We consider now the corrections to the $R^2$ coupling.
In the case of the model without massless states on the currents, 
the perturbative heterotic expression reads (Ref. \cite{gkp2}, Eq.~(4.25)):
\be
{16 \, \pi^2 \over g^2} = 16 \, \pi^2 \, S_2 -2 \log T_2 |\theta (T)|^4 
-2 \log U_2 |\theta (U)|^4 \, ,
\ee
where for simplicity we omitted to specify the infrared, cut-off dependent 
running. The first term on the r.h.s., dependent 
on the field $S$, is replaced on the type~IIA side by an expression 
analogous to the other two terms. From Eq.~(\ref{000}),
we see in fact that, after the appropriate substitutions of fields,
the type~IIA correction reads:
\be
{16 \, \pi^2 \over g^2} = -2 \log S_2 |\theta_2 (S)|^4 
-2 \log T_2 |\theta (T)|^4  -2 \log U_2 |\theta (U)|^4 \, ,
\label{r200}
\ee
The behavior for large-$S$ reproduces the heterotic tree level term $S_2$,
confirming compatibility with string--string duality.

In the freely acting heterotic orbifold 
with a perturbatively massless currents sector, instead, 
the perturbative correction doesn't have
a simple expression, with the contribution of the two
moduli $T$ and $U$ factorized in two terms.
This may signal that these moduli are not properly ``diagonalized'',
as is suggested also by the fact that,
had we computed the corrections to the effective couplings
of $F^2$ terms, we would have found that only in the case
of the orbifold without gauge group these are, trivially, given
by their tree level expression. In the other cases, non-zero
gauge beta functions lead in general to one-loop corrections of the tree level
term. This means that the ``true'' dilaton, ``$S^0$'',
is a function of the fields $T$ and $U$ (and of the various Wilson lines). 
This sounds quite reasonable: a shift on $S^0$, produced on
the type~IIA side by the projection $Z_2^{(1)}$, would therefore
appear as a shift not only on the field $S$ but also on $T$ and $U$,
leading to an observable twist of the heterotic currents.

\subsubsection{\it the Type I dual}

Considerations analogous to those we made for the heterotic dual
apply also for the type~I dual of this construction. Here, 
not only the duals of the three 
hypermultiplet moduli $U^{(i)}$ are invisible, either because twisted or 
because missing, but also two out of the three $T^{(i)}$: 
they must correspond in fact
to the two coupling-constant fields $S^{\rm I}$ and $S^{\prime}$.
As a consequence, two of the three mass expressions given in (\ref{m123})
correspond to a non-perturbative mechanism.
From the type~I point of view, the masses of the states of the three
type~IIA twisted sectors, Eq.~(\ref{m123}), read:
\ba
m^2_{(1)} & \sim & {1 \over S^{\rm I}_2} \, , \nonumber \\ 
m^2_{(2)} & \sim & {1 \over S^{\prime}_2} \, , \label{m123I} \\ 
m^2_{(3)} & \sim & {1 \over U_2} \, . \nonumber  
\ea 
For what we saw, one of the above three expressions
always bears a ``hidden'' dependence on one of the other two moduli,
so that the actual, ``physical'' mass eigenvalues are not ``orthogonal''.
In the case in which the masses of all the three twisted sectors
of the type~IIA model have a perturbative type~I counterpart,
namely, when $m_{(1)}$ and $m_{(2)}$ depend also on $U$,
the dual is an orientifold without open string sector, 
obtained by applying a freely acting $Z_2$ projection,
like the one of the ``momentum breaking'' model, to the orientifold 
described in Section \ref{sbroken}. The D9-branes sector is
therefore absent because of the free action of $\Omega^{\cal F}$,
and, owing to the free action of the further $Z_2$ projection,
there are no D5-branes as well. Supersymmetry appears on the other
hand to be spontaneously broken, as required by string--string duality.

However, as for the case of the heterotic dual, even here the
mass dependence of the two potentially perturbative sectors     
of the type~I string, namely the D9- and the D5-branes sectors,
may appear completely non-perturbative. In this case,
depending on whether the moduli $S$ or $S^{\prime}$ are large or small,
we fall into a situation described either by a ``winding breaking'',
or by the ``momentum breaking'' model presented in Ref. \cite{adds}. 
In any case, ${\cal N}=2$ orientifolds in which a super-Higgs mechanism 
produced by a freely acting $Z_2$ orbifold projection
lifts the mass of one or two sectors. For the reasons
already discussed in the previous section, we argue that,
although a priori possible, the ``winding breaking'' case
has to be excluded, being most probably the dual of the ``finite''
theory of section \ref{19,19}.

The ``momentum breaking'' situation is on the other hand dual
to the $Z_2$ freely acting heterotic orbifold with 
perturbative gauge group originating from the currents, that we have
discussed in the previous paragraph.
Although, out of the Abelian point, the gauge couplings receive
perturbative corrections that, on the heterotic side, read:
\be
{1 \over g^2} \sim S_2 + \beta_1 \log T_2 | \theta (T)|^4 
+ \beta_2 \log U_2 | \theta (U)|^4 + 
{\rm e}^{-(T,U)} + {\rm n.p.} \, ,
\ee
nevertheless duality with the type~I construction is in this case 
compatible with the presence of a non-zero $\beta_1$. 
The contribution of the field $T$, dual to
the type~I field $S^{\prime}$, is in fact entirely non-perturbative
from the type~I point of view. Only in the limit in which this term 
diverges linearly we would get an apparent contradiction, because this
term would reproduce the tree level contribution of a D5-branes sector.
But indeed, in such a limit, the action of the orbifold projection
is not anymore free, and most probably, as argued in Ref. \cite{gk},
there appears a D5-branes sector.

To {\bf summarize}, the field/sector identifications of these
constructions read:
\begin{table}[here]
\begin{center}
\begin{tabular}{c | c | c | c | c | c |}
& {\rm Het.} & {\rm Type~I} & {\rm Type~IIA} & {\rm rank} & $R^2$ 
{\rm behavior } \\ \hline
{\rm sector/coupling} & $S$ & $S^{\rm I}$ & $T^{(1)}$ &
{\bf 0} & $\theta$ \\ \hline
{\rm sector/coupling} & $T$ & $S^{\prime \, {\rm I}}$  
& $T^{(2)}$ & {\bf 0} & $\theta$ \\ \hline
{\rm sector/coupling} & $U$ & $U^{\rm I}$
& $T^{(3)}$ & {\bf 0} & $\theta$ \\ \hline
\end{tabular}
\end{center}
\label{table3,3}
\caption{The duality identifications for the "$(3,3)$" vacuum.}
\end{table}

\noindent
As discussed, we argue that, besides the explicit construction of 
Ref. \cite{gkp2}, reported in section \ref{htd}, also
all the freely acting heterotic orbifolds in which the
(perturbative) gauge group is realized at the level one 
fall in this class: all the states of the currents are non-perturbatively
lifted to non-zero mass. The same is true for the ${\cal N}_4=2$
freely acting orbifolds of the ${\cal N}_4=4$ type I string 
obtained as "momentum" breaking, not to be confused with the
"freely acting orientifold"  of section \ref{sbroken}.
The difference between the explicit construction "without"
gauge group and those with an apparent gauge group would consist
only in an inversion of the fields $S$ or $S^{\rm I}$,
respectively.

\vspace{.7cm}

\subsection{\sl The M-theory point of view}
\label{mpw}

\vspace{.5cm}

In the previous sections we have seen some examples of type~II\big/heterotic
dual pairs. We have also argued that in some cases, the perturbative heterotic
gauge group, together with all the states originating from the currents,
are non-perturbatively lifted to a non-zero mass.
More specifically, we have argued that this is the case for the
heterotic constructions that are not in the moduli space of 
constructions with a symmetric embedding of the 24 instantons in the two
$E_8$ factors, as is instead the case of the $U(16)$ model \cite{6danom}.
We have argued that this is the case also for the ``freely acting''
orbifolds, namely orbifolds in which the breaking of supersymmetry
appears to be ``moduli dependent'', with the exception of some very particular
cases. These latter correspond to ``finite'' theories, in which a twist in the 
currents produces an $N_V=N_H$, reduced rank realization of the gauge sector,
in which vectors and hypermultiplets arrange into multiplets of an
effective ${\cal N}_4=4$ supersymmetry.
Here we try to justify our hypothesis by considering the situation
from the M-theory point of view. We consider first the cases in which 
the reduction of supersymmetry is not spontaneous, and then the case
of freely acting projections.

As is known since the appearance of Refs. \cite{hw1,hw2},
the (ten dimensional) $E_8 \times E_8$ heterotic string is obtained by
orbifolding the M-theory by a $Z_2$ reflection along the eleventh 
coordinate. The ``twisted sector'' of this orbifold is represented
by the states of the two $E_8$ factors of the gauge group, ``sitting''
each one at one of the two fixed points.
Therefore, the existence of a (massless) gauge sector, besides the
pure reduction of the ${\cal N}_{11}=1 $ supergravity multiplet,
is related to the existence of fixed points in this orbifold. 
Had we performed instead a freely acting projection, such as
a Scherk-Schwarz projection of M-theory,
we would have obtained a massless spectrum constituted by the 
simple reduction of the supergravity multiplet.
When we say that the eleventh coordinate is twisted, we mean that
we are in a ``rigid'' situation, in which a continuous motion
from a fixed point to the other is not allowed, and the theory
doesn't distinguish between different values of the radius of this coordinate.
This situation must be compared to the case of a Scherk--Schwarz projection,
in which such a motion instead makes sense.
Only after compactification to lower dimensions, thanks to the
embedding of the previous one-dimensional problem into a higher
dimensional geometric space, we have continuous parameters
that allow a displacement from a fixed point to the other. This is nothing but a rephrasing of the well
known fact that in lower dimensions we can introduce Wilson lines
and break differently the two factors of the gauge group 
\footnote{A Wilson line allows to distinguish the two fixed points.
It is worthwhile to see how this works.
Let's consider the 9-dimensional case, where we have introduced a 
Wilson line that breaks one of the two 
$E_8$ to a subgroup. The effective coupling
of this group is different from the coupling of the other $E_8$:
now it depends not only on the dilaton of 9 dimensions but also on
the new field parameterizing the Wilson line.
Let's now reduce the size of the eleventh dimension, so as to
bring one fixed point of the Horava--Witten orbifold close to the other one.
If we don't change  the radius of the 9-th coordinate,
we reduce also the effective coupling of 9 dimensions, and in the limit
in which the two fixed points coincide, we have actually a free
theory. Therefore, in order to get a meaningful reduction, 
we must also change the radius of the 9-th coordinate, 
and decompactify the theory
in order to keep fixed the effective coupling. But then
we recover an effectively ten dimensional theory, and the effect
of the Wilson line has disappeared. Therefore, as the two fixed points
get closer and closer to each other, the theory comes back to the initial one,
in which the gauge groups on the two fixed points are equal.}.

\subsubsection{\it "rank 48": the symmetric embedding}

We now reduce further the amount of supersymmetry, by
applying a second $Z_2$ orbifold projection. This acts as a twist
on coordinates different from the eleventh one. Therefore,
we twist also compact coordinates that previously 
we used in order to ``move'' the theory away from the rigid initial
condition. This means that, unless we further compactify the theory
and introduce Wilson lines in the extra compact coordinates,
when introducing the new orbifold projection
we must respect the twist symmetry of the Horava--Witten orbifold.
Namely, the embedding of the spin connection into the gauge group
must be done symmetrically in the two $E_8$.
This argument is intriguingly related to those that, in Ref. \cite{dmw}, 
led to argue that, in six dimensions,
the ``(12,12)'' construction possesses an S-duality
symmetry. As discussed in Ref. \cite{dmw}, owing to the embedding
of the spin connection into the gauge group, 
this construction is more complex than in a simple $Z_2 \times Z_2$ 
orbifold. However, as far as only the fundamental
operation underlying the construction and the coupling are concerned,
with the Wilson lines and the choice of the gauge bundle treated as
``second order'' corrections, one is already guided to a correct answer
by simply considering the symmetries of the orbifold.
In this case, by considering the construction from the
dual type~II point of view, we see that there must be  a deep relation between 
S-duality and masslessness of the perturbative sector of the heterotic string.
If we see this sector as the twisted sector of an orbifold (e.g. 
Horava--Witten theory or type~IIA string), whose coupling is parameterized by
$S$, the relation is the same 
as the one between any twisted sector of an orbifold and the $SL(2,Z)$
symmetry of the modulus parameterizing the corresponding coupling
constant. This is the reason why we expect the $U(16)$ heterotic
orbifold to be dual to the CY$^{51,3}$, $Z_2 \times Z_2$ orbifold
of the type~IIA string.

\subsubsection{\it CY$^{19,19}$: the asymmetric embedding}

An asymmetric embedding is allowed only when the eleventh coordinate
is not rigidly twisted. But this means that in this case,
as in a Scherk-Schwarz projection,
there is no massless sector corresponding to the Horava--Witten projection. 
In the light of this discussion, it is perhaps not too surprising
that the only examples of heterotic constructions of 
Refs.~\cite{kv,kklm} for which it has been given a strong 
evidence of duality with the type~IIA string, belong to the class of the
symmetric embeddings. Since in fact the heterotic dilaton is mapped into
a perturbative modulus of the type~IIA string, what on the heterotic
side appears as a non-perturbative lifting of the mass of the gauge sector,
on the type~IIA side would be a completely perturbative phenomenon. Therefore,
we would expect to not find in the compact manifold (K3 fibration) the 
cycles corresponding to these states.
Our arguments are supported, at least naively, by the observation that,
for asymmetric embeddings, S-duality is broken \cite{dmw}.
The type~II orbifold realization of this situation is provided by
the CY$^{19,19}$ construction of section \ref{19,19}.

\subsubsection{\it freely acting orbifolds: CY$^{11,11}$ and CY$^{3,3}$}

The $Z_2$ freely acting orbifolds are constructed by coupling the $Z_2$ twist
on four coordinates to a shift on a further coordinate (therefore, they cannot
be constructed, at least perturbatively, in six dimensions).
Let's consider the situation from the heterotic point of view, i.e.
from the point of view of the Horava--Witten orbifold of M-theory.
As is known from the analysis of Ref. \cite{adk}, 
from heterotic/type~II duality in less than six dimensions
one derives that a \emph{perturbative} shift along a circle of the
compact space of the heterotic string is \emph{always} accompanied
by a ``shift'' also along the coupling constant \footnote{Notice that
here we are not talking about the coupling of the ten dimensions,
the eleventh coordinate of M-theory, but of the lower dimensional
coupling, of the five or four-dimensional theory.}.
It always contains therefore a freely acting projection  
on the dilaton field. From the M-theory point of view,
this corresponds to a Scherk--Schwarz mechanism, that ``moves''
the fixed points, lifting the mass of the perturbative heterotic
gauge group. This is the reason why we argue that in all the 
$Z_2$ freely acting orbifolds, such as those considered in 
Ref. \cite{kkprnew}, the states originating from the currents
are all non-perturbatively massive; from the Type IIA point of view,
this corresponds to the CY$^{3,3}$ orbifold.
The only way of preserving the masslessness of certain states in this
sector is by coupling the freely acting $Z_2$ orbifold projection to
another freely acting projection, that picks a shift in the same
direction, so that the two shifts add up to zero~\footnote{i.e.
to an integer shift, equivalent to a simple relabeling of the
quantum numbers.}. This is what happens in the 
orbifold realization of the CY$^{11,11}$ model, where,
besides the freely acting, supersymmetry breaking reflection,
it acts also, at the same time, a freely acting reflection
along the eleventh coordinate of M-theory. This operation 
exchanges the two $E_8$, and produces a level two gauge group, 
by lifting the mass of the
off-diagonal states by a shift along the ``dilaton coordinate''.
The two shifts along this coordinate therefore add up to zero,
and the whole operation doesn't contain any shift on the dilaton,
on which it acts as an ordinary, non freely-acting orbifold.
There exist therefore massless twisted states of the Horava--Witten orbifold;
they are identified two by two and give rise to a level two
realization of the currents.

\vspace{.7cm}

\subsection{\sl Reduced rank}
\label{reducedrank}

\vspace{.5cm}

In the type~II string there exists only one operation,
compatible with the $Z_2$ orbifold projections, that allows to reduce the
number of orbifold fixed points, and therefore, in general, 
also the number of $U(1)$ vector multiplets in the orbifold twisted sectors.
This is the ``$D$'' projection introduced in Ref. \cite{dvv}
and applied in Refs. \cite{6auth,gkp,gkr}, in the framework of
type~IIA/heterotic duality. As a pure ``geometrical''
projection, this operation makes in fact sense for any $Z_2$ orbifold,
whether of the type~II, or the heterotic, or even the type~I string.
By considering appropriate configurations of projections of this kind,
it has then been possible to construct the heterotic duals of the
type~II orbifolds corresponding to CY$^{7,7}$ and CY$^{5,5}$ \cite{gkp}.
Their type~I dual can be straightforwardly constructed in a
similar way, by simply $D$-projecting the type~I dual of the
CY$^{11,11}$ model of section \ref{11,11}.

On the type~I side, however, the rank of the gauge group can be reduced also
by introducing a quantized $B$ field \cite{b,ang,kakush}.
In this section we concentrate on this construction, and show that,
from the heterotic and type~II point of view, this too can be
interpreted in terms of $D$-projections.

\subsubsection{\it on the type~I side: the $B$ field}

At the ${\cal N}_4=4$ level,
type~I strings with a reduced rank gauge group \cite{b}
are perturbatively dual to analogous heterotic 
constructions. Although natural this may seem, the fact that duality 
with the heterotic string works is not so obvious. The $B$ field 
couples in fact to the windings,
as it can be seen from the formula for the left and right moving
momenta \footnote{We recall
that the orientifold projection identifies, modulo integers, 
$p_{\rm L}$ with $-p_{\rm R}$.}:
\be
p^i_{\rm L(R)} =m^i \pm {1 \over \alpha^{\prime}} \left( g_{ij} \pm
B_{ij} \right)n^j \, .
\label{bij}
\ee
The introduction of a non-vanishing, quantized $B$ field implies,
on the type~I side, an action on the windings. 
It is not therefore an operation on the Kaluza--Klein sector of the theory,
something that, by adiabaticity, would ensure us to find a corresponding
operation, with the same effect, on the heterotic side.
Indeed, duality works because the ${\cal N}_4=4$ supersymmetric
theory is very constrained, and the heterotic
string with ${\cal N}_4=4$ and gauge group of rank 16 
does not contain non-perturbative massless sectors and possesses
S-duality. It does not really matter therefore whether the
operation we are doing on the theory is visible at the weak coupling or not.
These ambiguities are however unraveled when we instead consider the
theory with ${\cal N}_4=2$ supersymmetry.
The introduction of a $B$ field in the $U(16) \times U(16)$ model
halves the rank of the gauge group in both the D9- and D5-branes sectors,
reducing the gauge group to $U(8)_9 \times U(8)_5$ (see Ref. \cite{ang}).
In order to see what is the heterotic dual, we must proceed as in Section
\ref{19,19}: we must convert the operation on the windings
into an operation on the momenta, and there translate into
heterotic fields. If we do that, we see that both the D9- and D5- branes
sectors on which the $B$ field has reduced the rank of the gauge group
correspond to non-perturbative sectors.

\subsubsection{\it on the heterotic side}

From the heterotic point of view, the dual of the type~I $B$ field is 
a projection that ``moves'' half of the
orbifold fixed points. On the twisted coordinates, it acts therefore as a
$Z_2$ exchange of twist fields: $\sigma_+ \to \sigma_-$ \cite{dvv}, 
while on the untwisted two-torus  it acts as a $Z_2$ translation:
it is therefore a ``$D$-projection'' 
\footnote{For more details about this operation, we refer the reader to
Refs. \cite{gkr,6auth}.}.
When applied to the dual of the ``$U(16)$'' model,
the visible effect of this projection is only that of halving
the number of hypermultiplets originating from the twisted
sector. This already goes in the correct direction, because we
know that these are dual to the type~I hypermultiplets in the 
twisted closed sector, and therefore in one-to-one correspondence with those
in the bi-fundamental representation of the D9- and D5-branes gauge groups.
The corrections to the $R^2$ term read (see Ref. \cite{gmonobis},
Eqs.~(5.14) and (5.17)):
\be
{16 \, \pi^2 \over g^2} \; \sim \; 16 \, \pi^2 \,S_2 \,
- 4 \log T_2 \left( {3 \over 2} | \eta (T)|^4 \, +
{1 \over 2} | \theta (T)|^4 \right) \, 
-4 \log U_2 \left( {3 \over 2} | \eta (U)|^4+
{1 \over 2} | \theta (U)|^4 \right) \, .
\ee
The particular dependence on the fields $T$ and $U$, with a 
combination of eta and theta functions, is a signal of the 
rank reduction in these sectors. The fields $T$ and $U$ parameterize
in fact the coupling constants of these sectors, and
the linear behavior for large-$T$ and/or large-$U$, that should account
for the inverse of their ``bare'' square couplings, is 
one-half of the one without rank reduction, as expected for ``level
two'' realizations of the gauge algebra.
This supports the conjecture that these fields are dual to the
fields $S^{\rm I}$ and $S^{\prime}$ of the type~I side, where the analogous
correction reads \footnote{The expression is immediately derived by using 
the techniques of Ref. \cite{gk}, from the partition function of the model, 
quoted in Ref. \cite{ang}. It is basically the same as (\ref{nonsoI}),
apart from a factor 1/2 of rescaling of the one-loop contribution, due to the
projection introduced by the $B$ field. This can be understood by considering
that the $B$ field in practice "halves" the model in the non-trivial
orientifold sectors. The regularized expression of the $R^2$ amplitude
involves therefore a series of terms all divided by a factor 2.}:
\be
{16 \, \pi^2  \over g^2} \; \sim \; 16 \, \pi^2 \, S_2 \; + \; 
16 \, \pi^2 \, S^{\prime}_2 \; 
- 6 \log U_2 \vert \eta ( U^{\rm I} ) \vert^4 \, .   
\ee
As in section \ref{rank48}, it is the field 
$U^{\rm I} \equiv U^{\rm type \, I}$ which
is dual to the heterotic dilaton--axion field $S^{\rm het}$. 
The map between the heterotic moduli $T$ and $U$ and the couplings
of the type~I dual is in this case: $T \leftrightarrow 2 \tau^{\rm I}_S$,
$U \leftrightarrow 2 \tau_{S^{\prime}}$, with the typical factor 2
in agreement with the level doubling.

\subsubsection{\it on the type~IIA side}

From the type~IIA point of view, this construction corresponds
to a specific operation performed on the fiber of the space
described in Section \ref{rank48}.
As we said, at the point in which the fiber degenerates, there appear
two other K3 fibration structures, whose bases, corresponding to
the cycles $T$ and $U$, are inside the fiber of the ``original''
fibration, based on $S$. The type~IIA dual of the projection
induced by the quantized $B$ field 
is therefore a projection that halves the volume of the ``original'' fiber
($ \equiv $ half number of orbifold fixed points 
on the heterotic side), halving thereby the volumes of the bases of 
the two new fibrations, represented by $T$ and $U$. 
As discussed also in Ref. \cite{hmFHSV},
such an operation halves the rank of the respective gauge groups.

The explicit dual of the type~I string with group $U(8)_9 \times U(8)_5$
\cite{ang} was constructed in Ref. \cite{gkr},
and corresponds to a $Z_2 \times Z_2$ orbifold with Hodge numbers (31,7),
obtained from the CY$^{51,3}$ by applying a $D$-projection to two 
twisted sectors. According to Ref. \cite{gkr}, Eqs.~(E.24), (E.27),
the correction to the effective coupling of the $R^2$ term reads:
\ba
{16 \, \pi^2  \over g^2} & = & 
- 2 \log T^{(1)}_2 \left( {3 \over 2} | \eta (T^{(1)})|^4 +
{1 \over 2} | \theta (T^{(1)})|^4 \right) \, 
-2 \log T^{(2)}_2 \left( {3 \over 2} | \eta ( T^{(2)})|^4+
{1 \over 2} | \theta (T^{(2)})|^4 \right) \nn \\
&& -6 \log T^{(3)}_2 \vert \eta ( T^{(3)} )  \vert^4 \, ,
\ea
suggesting the correspondence of the fields $T^{(1)}$, $T^{(2)}$, $T^{(3)}$
respectively with the heterotic fields $T$, $U$, $S$ and with the type~I
fields $S^{\rm I}$, $S^{\prime}$, $U$.
Two twisted sectors have indeed a reduced rank, with 8 vector multiplets
in each, and no hypermultiplets, as expected from the rank reduction of
the orbifold of Section \ref{51,3} in two twisted sectors.
Surprisingly, the sector that should correspond to the perturbative
heterotic sector, that we would expect to remain untouched,
contains now only 12 vector multiplets, and four hypermultiplets.
In order to understand this apparent discrepancy, we must keep
in mind that, on the type~IIA side, the spectra that appear
at the orbifold point contain only the states that can be explicitly
constructed with vertex operators of the world-sheet conformal theory,
and they must, by construction, respect the factorization of the 
target compact space into the three tori, produced by the
$Z_2 \times Z_2$ projection. Therefore, as we remarked in section \ref{51,3}, 
not only the  states charged under the gauge group don't appear 
at the orbifold point
(they are non-perturbative, D-branes states), 
but also the states that are ``multi-charged'', i.e. charged under 
the groups of two different twisted sectors.
In the orbifold (51,3) of Section \ref{51,3}, each twisted
sector contains 16 vector multiplets.
As we discussed, they indeed correspond to the ``projection'', in the
sense of above, of the $U(16)$ gauge group.
Let's concentrate on the two $U(16)$ visible on the
type~I side, corresponding to the $T$ and $U$ sectors: 
the charged hypermultiplets are in
the $({\bf 120},{\bf 1}$, $({\bf \overline{120}},{\bf 1})$,
$({\bf 1},{\bf 120})$, $({\bf 1},{\bf \overline{120}})$ and
in the $({\bf 16},{\bf 16})$. 
Let's now halve the rank of the gauge group in both the 
``$T$'' and ``$U$'' sectors, thereby producing,
as maximal group, $U(8)$ in each of them. This affects also
the sector which is perturbative on the heterotic side: since 
the number of twisted hypermultiplets
has been halved, the one-loop gauge beta
function is not anymore zero. But this is not compatible 
with a type~IIA orbifold. It would work if also the gauge group in the
``$S$'' sector had been reduced to $U(8)$, with 56 hypermultiplets
in the $(\bf{28}, \bf{1})$ and $(\bf{1},\bf{28})$ and 64 as
$8 \times \bf{8}$ of $U(8)$. This however corresponds to a ``too big'' 
reduction \footnote{It in fact exists,
and is found by further reducing the model also in the ``$S$'' sector
(see Ref. \cite{gkr}, where it appears with 8 vector multiplets in each
of the three twisted sectors).}. The structure of the spectrum of the dual
sector on the type~IIA side seems rather to correspond to  a group 
$U(8) \times G|_2$, where $G|_2$ is a gauge group of rank 4, with vector 
and hypermultiplets in isomorphic representations, $G_V \cong G_H$, 
and the algebra realized at the level 2, as in section \ref{11,11}. 
This can be for instance $\left( SU(2)\vert_2 \right)^4$, 
the maximal allowed being $G_V \cong G_H = SO(8)$.
Both $U(8)$ and $G|_2$ have vanishing beta  
functions, and the group is the one with the maximal rank compatible
with the type~IIA conformal theory. With a non-vanishing gauge beta function, 
there would have been a mixing of fields, signaling that
$S$, $T$, $U$ were not properly diagonalized as are $T^{(1)}$,
$T^{(2)}$, $T^{(3)}$. In such cases, as we discussed at the end of 
Section \ref{3,3}, the breaking of the $SL(2,Z)$ symmetry
of the coupling field of one sector, with its related
lifting in the mass of the states, affects also other sectors.
When the breaking of this symmetry is due to the supersymmetry breaking
projection, it results, as we have seen, in the lifting of all the 
states of the corresponding sector. When instead it is due to a 
``rank reducing'' projection, it involves only a part of the states.
In this specific case, as a consequence of the reduction in
the other sectors, there would have been a partial reduction
also on the ``$S$'' sector. This is precisely what we argue it happens
on the heterotic side: since the beta
function is not vanishing, the field $S$ is not ``orthogonal'' to the
fields $T$ and $U$. As a consequence, there is a partial,
non-perturbative reduction of the spectrum, up to the maximal
group $U(8) \times G|_2$.
As a check that this is indeed possible, we make sure that
the states of this configuration were already contained in the
spectrum of the $U(16)$ model. This is not a problem for the 
$U(8)$ factor, for which it is obvious.
For the $G|_2$ with hypermultiplets in an isomorphic
representation, we just observe that the maximal group allowed
is $SO(8)$. The hypermultiplets can originate from those in the
$\bf{120}$ of $U(16)$, equivalent to the Adjoint of $SO(16)$,
now broken to $SO(8)$ \footnote{A further breaking
is to $SO(4) \times SO(4) \cong SU(2)^4$.}. 
We remark that such a ``reduction'' in this sector is also
required by consistency with the interpretation we gave in section
\ref{3,51} of the type~I hypermultiplets
originating from the twisted closed string sector of the orbifold.
The introduction of a $B$ field apparently does not affect the twisted 
closed sector. On the other hand, we observed that the 
hypermultiplets in this sector are ``paired'' to those originating from 
the D5-branes, and argued that they are identified with the perturbative 
hypermultiplets of the heterotic string.
Therefore, a rank reduction on the D5-branes must reflect in a similar
reduction on these states, that we interpreted as the ``projection'' on the
Cartan of a ``hypermultiplet'' group. This reduction must be the same
happening on the heterotic side. In this case, the reduction is from
$U(16)$ to $U(8)$. The remaining hypermultiplets, transforming in the $G|_2$,
are not supposed to be ``paired'' to those of the D5-branes.
This accounts for the apparent ``discrepancy'' between twisted closed
and open sectors of the type~I string with a $B$ field.

As usual, we {\bf summarize} the results of our analysis in a table:
\begin{table}[here]
\begin{center}
\begin{tabular}{c | c | c | c | c | c |}
& {\rm Het.} & {\rm Type~I} & {\rm Type~IIA} & {\rm rank} & $R^2$
{\rm behavior } \\ \hline
{\rm sector/coupling} & ${\rm S}$ & ${\rm U}^{\rm I}$ & ${\rm T}^{(3)}$ &
{\bf 8} $\oplus$ {\bf 4}$|_2$ & $\eta$ \\ \hline
{\rm sector/coupling} & ${\rm T}$ & ${\rm S}^{\rm I}$ & ${\rm T}^{(1)}$ &
{\bf 8} & $\eta + \theta$ \\ \hline
{\rm sector/coupling} & ${\rm U}$ & ${\rm S}^{\prime \, {\rm I}}$ 
& ${\rm T}^{(2)}$ & {\bf 8} & $\eta + \theta$ \\ \hline
\end{tabular}
\end{center}
\caption{The duality identifications for the simplest reduced rank vacuum.} 
\label{rankr}
\end{table}

\noindent
Notice the relation between $R^2$ behavior and the rank of gauge group
on the various sectors: as it is clear also
from all the other tables, $\eta$ always 
corresponds to a maximal rank realization
(either ${\bf 16}$ or ${\bf 8}|_2$, or ${\bf 8} \oplus {\bf 4}|_2$). 
$\theta$ signals the "absence" of states, 
because lifted to a non-vanishing mass. Mixed cases, in which the 
reduction is only partial, correspond to a mixed behavior also of the
threshold correction.

\vspace{.7cm}

\subsection{\sl The complete classification}
\label{complete}

\vspace{.5cm}

If we now consider the complete classification of the $Z_2 \times Z_2$
symmetric orbifolds of the type~II string,
summarized in table D.1 of Ref. \cite{gkr}, we see that
all the orbifolds are obtained from the five cases we
have analyzed in this chapter, by simply applying rank-reducing,
semi-freely acting projections of the above described type.
Therefore, the analysis of all the remaining cases can be performed
by straightforwardly combining the rules we have presented in the
previous sections. For instance, the mirror
of the CY$^{31,7}$ model described in section \ref{reducedrank}
corresponds to a type~II orbifold in which the massless states
of the three twisted sectors, namely 8 hypermultiplets per each
in two twisted sectors, 12 = 8 + 4 hypermultiplets and 4 vector multiplets
in the third twisted sector, have to be interpreted as follows.
The theory has two sector with gauge group $U(8)$ and,
besides hypermultiplets in the $8 \times {\bf 8}$, also hypermultiplets
in the ${\bf 28}$ and ${\bf \overline{28}}$, that can be interpreted 
as the Adjoint of $SO(8) \times SO(8)$. In the CY$^{31,7}$ orbifold we see,
in each of these two sectors, only the eight vector multiplets of the 
Cartan of $U(8)$; in the mirror, we see only the hypermultiplets in
the Cartan of $SO(8) \times SO(8)$.
In the third twisted sector, an analogous exchange of role involves
the $U(8)$ part, while the rank 4, $G|_2$ part is self-mirror: vector
and hypermultiplets appear there in isomorphic representations.

In a similar way, one can perform the analysis for all the cases of table
D.1 of Ref.~\cite{gkr}, completing the analysis of the 
type~II\big/heterotic\big/type~I duality for all these orbifolds. 
Among these, we notice the relevant cases in which rank reducing projections
act on the sector corresponding to the field $S$, namely the
heterotic currents. This operation is entirely non-perturbative
from the point of view of the heterotic string, where it is not
observable. An example is provided by the type~II orbifold 
corresponding to CY$^{24,0}$ (and by its mirror CY$^{0,24}$). In general,
whenever discrepancies in the spectra are found, one has always to
check whether these are due to a real difference of the theories,
or to simple technical facts, keeping in mind  that:

\begin{itemize} 

\item in the type~II string, as well as in the closed sector of the type~I 
string, the twisted fields associated to the orbifold fixed points,
no matter or whether vector multiplets or hyper-multiplets,
can only account for states transforming in an Abelian group.
Non-Abelian extensions are therefore non-perturbative, of which
only the Cartan subgroup is visible;

\item in the type~II string there can never appear bi-charged states, 
i.e. states charged under two different twisted sectors;

\item in the heterotic string the situation is reversed: owing to the
embedding of the spin connection into the gauge group,
the states associated to the orbifold fixed points are naturally
charged under both the perturbative gauge group and an extra,
non-perturbative gauge group. In this case, only bi-charged states
may appear, and therefore any extra gauge boson, associated
to special points of the K3 moduli, is non-perturbative.

\end{itemize}
After having selected which part of the spectrum 
is expected to appear on both sides,
the guidelines for the identification of the dual constructions
have to be found in the comparison of string amplitudes corrections.

\vspace{1.5cm}

\noindent

\section{\bf ${\bf \boldsymbol{\beta}_{\bf  gauge} \boldsymbol{\neq} 0}$: 
a puzzle of string theory}
\label{sbreaking}

In sections \ref{rank48}--\ref{complete} we have considered only
${\cal N}_4=2$ constructions with vanishing (one-loop) gauge beta 
functions.
We consider now heterotic orbifolds in which one or more factors 
of the gauge group have a non-vanishing beta-function. 
A non-vanishing beta function always introduces in  
the expression of the effective coupling a dependence on the moduli 
of the torus (and on the Wilson lines). Namely, the generic correction
is always of the form:
\be
{1 \over g^2} \; \approx \; S_2 \; + \; \Delta(T,U, Y^i) \, + \, {\rm n.p.} 
\, ,
\label{ggauge}
\ee
where $S_2$ is the universal, tree level contribution, given by the vacuum 
expectation of the dilaton, $\Delta(T,U, Y^i)$ encodes the one-loop
contribution, and ``n.p.'' stands for non-perturbative contributions,
of order ${\cal O}\left( {\rm e}^{-S_2} \right)$.
Owing to extended supersymmetry, the perturbative corrections to the
effective gauge coupling stop at the one loop.
For constructions in which the two-torus can be ``factorized'',
namely when the compactification is of the type  $T^2 \times K3$,
without any orbifold or Wilson line action on $T^2$,
the one-loop correction has an universal expansion for large
$T$ and/or $U$:
\be
\Delta_{T,U \gg 1} \; \approx \; \beta T_2 \, + \, \beta U_2 \, . 
\label{delta}
\ee
More in general, even when the torus is not factorized,
there is always a region in the moduli space in which
this function behaves as:
\be
\Delta_{X \gg 1} \; \approx \; X_2 \, ,
\ee
where $X$ stays for $T$, $U$, or $T^{-1}$, $U^{-1}$.
For what we saw in the previous sections, $X$ is always the 
coupling of a ``hidden'' sector. Its appearance in the expression
of the corrections of the effective gauge coupling is somehow
puzzling: the appearance of these hidden sectors is tuned
by the moduli of the K3, i.e. the moduli  in the hypermultiplets.
On the other hand, owing to ${\cal N}_4=2$ supersymmetry,
expression (\ref{ggauge}) can only depend on the moduli
in the vector multiplets.
How is it then possible that a gauge coupling of the
perturbative sector is affected by a coupling of a hidden sector,
in a way that apparently is not sensitive to the actual
existence of such a hidden sector? Intuitively, we would find
reasonable to expect that, as the hidden sector becomes massive,
its contribution to (\ref{ggauge}) is suppressed, and
the more and more suppressed as the mass of these states increases,
up to the limit in which they are so heavy that they ``decouple''.
At this point, we would expect that also the dependence of (\ref{ggauge})
on their coupling $X$ drops out.  However, the mass of these states depends 
on the moduli in the hypermultiplets. Therefore, it cannot
appear in (\ref{ggauge}). Our aim here is then to understand the meaning, 
and what are the implications, of the appearance of such a coupling, $X$, 
in the expression of an effective coupling of the perturbative sector.

Clearly, from a field theory point of view the appearance of this field 
in the renormalization of the gauge couplings can only by justified
by the running of states charged under both the sectors.
These states should belong to the perturbative spectrum,
and appear as associated to the orbifold fixed points, or, in general, 
to certain cycles of the compact space. However, as we have seen, 
such states seem to be
necessarily uncharged under the hidden gauge group, at least as long as
we give a description of the hidden sector in terms of elementary states
of a gauge field theory, as in the cases considered in the previous sections.
On the other hand, expression (\ref{delta})
is not quite the one we would find from a field theory analysis.
Let's in fact consider a generic effective field theory
with two gauge sectors, each one with its own bare coupling, $g$ and 
$g^{\prime}$, parameterized by a dilaton-like complex field, $S$
and $T$ respectively:
$1\big/ g^2 \equiv \Im S$, $1\big/ g^{\prime \, 2} \equiv \Im T$.
Let's also assume that, as in sections \ref{51,3}, \ref{3,51},
the degrees of freedom consist of: 
the vector multiplets of the two gauge sectors, each one with
only one (simple) gauge group, and a set of hypermultiplets charged under both
the gauge groups, $G$ and $G^{\prime}$. 
The leading contribution of $g^{\prime}$ to the correction of the coupling $g$
would be  at the  order ${\cal O} \left(g^{\prime \, 2} = 1 \big/ T_2 \right)$.
In particular, the one loop contribution
would come from diagrams like b) and c) of fig. \ref{wff}  
(page~\pageref{wff}, section~\ref{3,51}):
\be
{1 \over g^2} \;  \to  \; S_2 \, + \, b T_2^{-1} \, , 
\label{bbprime}
\ee
However, the coefficient $b$ would not account for
the entire beta-function coefficient: also diagrams
like b) or c), but without $W^{\prime}$
insertion, would give a non-vanishing contribution to the beta function of $g$,
that would not contain an order $g^{\prime \, 2}$. \newline
Moreover, the contribution of $T$ in (\ref{bbprime})
appears with the ``wrong'' power with respect to  the string corrections, 
expression (\ref{delta}). 

The point is that our diagram analysis has been necessarily 
performed at the weak coupling, $g,g^{\prime} < 1$, where, in order for
perturbation theory to make sense, the ``one-loop''
correction to a coupling, necessarily containing a positive power of the 
other coupling, must be ``small'', i.e. $< 1$. The string corrections,
instead, lead automatically to a one-loop correction like:
\be
\Delta \left( T \right) \; \approx \; T_2 \, + \, {\cal O} \left(\log T_2,
\exp - T_2  \right) \, ,
\ee
bound by an (approximate) $SL(2,Z)_T$ duality to be always greater than one.
This is not the expected behavior of
field theory perturbation. Rather, it looks like the behavior 
of the renormalization of a weakly coupled sector in ``contact'' with
a strongly coupled sector, with coupling: 
\be
g^{\prime} ~~ \sim \; \langle T_2 \rangle ~  > \; 1 \, .
\label{gt1}
\ee
If we include also the infrared cut-off running \cite{kk}, and for
simplicity we concentrate on the dependence on $T$, neglecting the
(analogous) dependence on $U$, the renormalization of the coupling,
as computed in the heterotic string, reads:
\be
{1 \over g^2} \; = \; S_2 \, + \, \Delta(T,U) \; 
\stackrel{T \gg 1}{\approx} \; S_2 \, + \, 
\beta T_2 \, - \, \beta \log \mu  ~~ + \ldots 
\label{beta}
\ee  
It is then clear that the running along a path of the renormalization group
can be equivalently seen as a redefinition of the ``bare''
coupling of the hidden sector, $T_2$, for fixed bare coupling of the
perturbative sector, $S_2$ \footnote{More precisely,
what is redefined is not really the field $T$ but the threshold
correction $\Delta(T)$, approximately behaving as $\Delta(T) \sim T_2$
for large $T$.}. However, the relation (\ref{gt1})
implies that now, if the effective coupling of the perturbative
sector decreases with the scale $\mu$, the coupling of the
hidden sector instead increases. There is no ``continuous''
communication between the situation (\ref{bbprime})
and (\ref{beta}): the $SL(2,Z)_T$ symmetry of the heterotic string tells
us that, once in the region (\ref{gt1}), the theory is ``confined'' to stay
at the strong hidden sector coupling.   
We don't know what is the mechanism that leads the theory in the hidden
sector to confine; we can however understand why, in order to switch on/off
a ``Wilson line'', leading to a non-vanishing beta-function,
the theory must be driven at the crossing line between the
two regions of parameters allowed by the $SL(2,Z)_T$ symmetry of the two-torus,
corresponding to the boundary between the $SL(2,Z)_T$ completions
of expressions (\ref{bbprime}) and (\ref{beta}). 
Let's in fact start from the situation ``with beta-function'',
Eq.~(\ref{beta}). If we want to switch off the Wilson line, and
restore a situation in which the beta-function vanishes, we must
first continuously deform the theory up to the point in which
the correction $\Delta(T)$ is at its minimum: if the Wilson line does not act
on the moduli $T$ and $U$, i.e. on the two-torus, this happens
precisely at the self-dual point $T=1$. Here it is possible for the theory
to jump from one region to the other \footnote{If instead we start 
from a model with vanishing beta-function, 
and generate a non-zero beta-function by introducing a Wilson line 
acting on the two-torus, there is in general a limit in which
$\Delta(X)$ vanishes, or behaves logarithmically. This behavior 
has to be interpreted, according to Refs. \cite{kk,solving}, as
``equivalent to zero''. In any case, switching on  Wilson lines
that generate a non-vanishing beta-function
automatically drives a hidden sector to the strong coupling.}.

\begin{figure}
\centerline{
\epsfxsize=5cm
\epsfbox{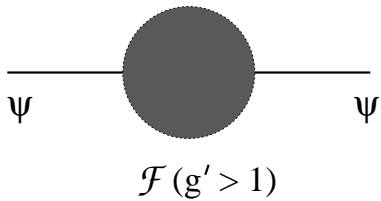}
}
\vspace{0.4cm}
\caption{the grey ball stands for a non-perturbative correction depending on
the strong coupling $g^{\prime}$.}
\label{ff}
\end{figure}

\begin{figure}
\centerline{
\epsfxsize=11cm
\epsfbox{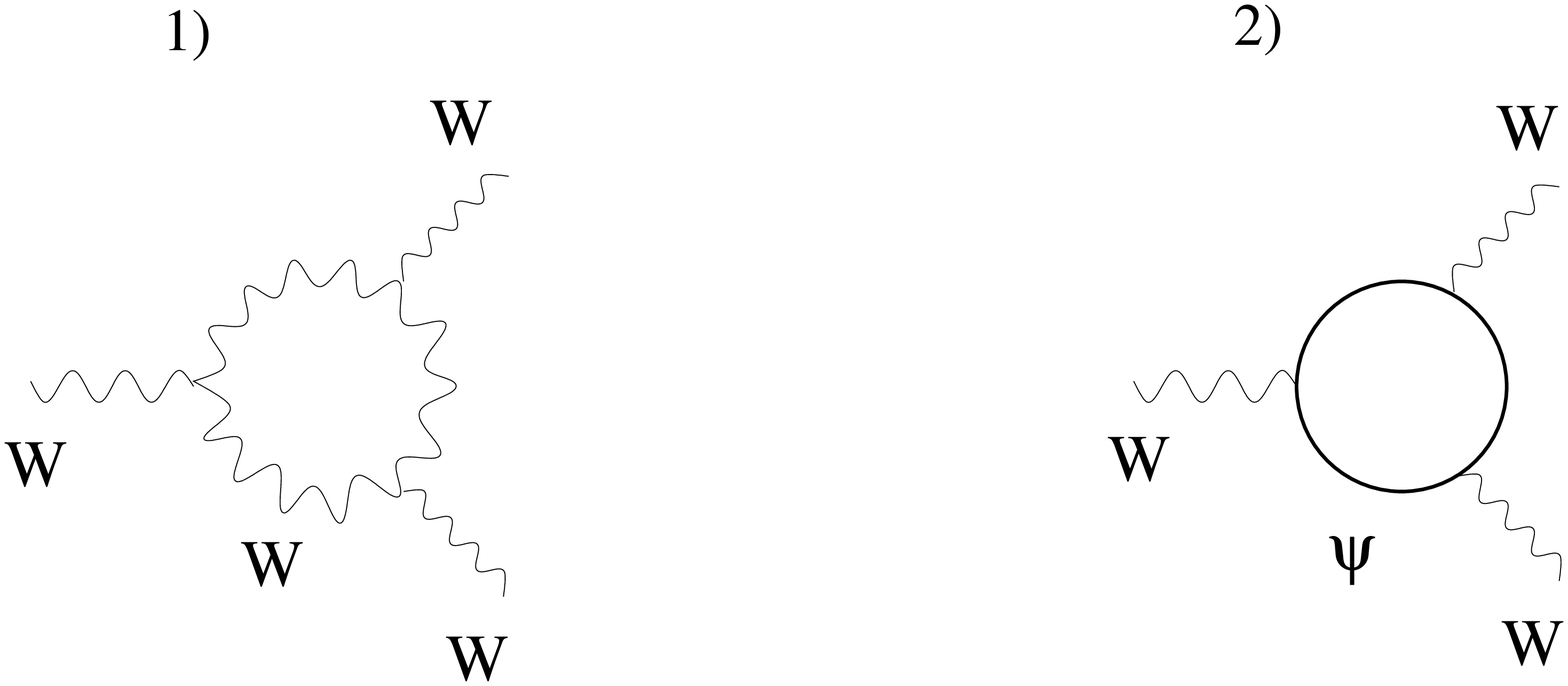}
}
\caption{the diagrams contributing to the one-loop beta function.}
\label{fbeta}
\end{figure}

The scenario appears  therefore to be the following: 
In order to switch on  Wilson lines leading to  non-vanishing gauge 
beta-functions, the hidden sector must first be brought out of the weak 
coupling ($g^{\prime} \sim 1$) region.
Then, after the ``Wilson line'' that generates a positive beta-function on 
the ``visible'' sector has been switched on,
the hidden sector is driven to the strong coupling $(g^{\prime} > 1)$.
There, a gaugino condensation \cite{dine,bdqq1,bdqq2} takes place, 
breaking supersymmetry.
In this phase, we cannot describe this sector in terms of 
elementary charged free fields. An analysis through such a kind of diagrams
is not anymore appropriate. Therefore, we cannot consider a correction
like the one corresponding to diagram a), based on a
weak coupling description of the hidden sector: it 
assumes in fact that the hypermultiplets are bi-charged in an elementary way.
As we have seen, this is not possible, and the $g^{\prime}$
dependence can only appear at the strong coupling, through corrections
of the type b), c) of figure~\ref{wff},
but due to renormalizations that not necessarily can be 
expressed as in the second line of that figure: these could well be
due to non-perturbative renormalizations of hypermultiplets propagators,
that we indicate in figure~\ref{ff}, where we cannot explicitly express 
the grey ball in terms of sums of usual vertices and propagators.
Only when they have a perturbative description in terms
of type~I string, the heterotic hidden sectors need in fact to appear
with a separation into charged sectors (on the D-branes) and
uncharged sectors (the closed string). We know on the other hand that,
in some way, the uncharged hypermultiplets of the closed string sector
``feel'' the D-branes, being, as the latter, in one-to-one correspondence with
the orbifold fixed points: it is therefore not too surprising that this effect
is emphasized at the strong coupling. 
We may ask how is it possible that a gauge group that in a weak coupling 
string description has a positive beta function,
at the strong coupling has instead an opposite behavior under renormalization.
The positiveness of the beta function  
is due to a predominancy of the hypermultiplets contribution, through 
diagrams of the type 2), with respect to the vector
contribution, given by diagrams of the type 1) of figure~\ref{fbeta}.
%\vspace{0.8cm}
%\noindent
At the strong coupling, the asymptotic states in the ``hypermultiplets''
are on the other hand forced to bound into singlets of the gauge group. 
The ``elementary'' vertices of bosons and matter states, whose coupling is the
charge of the matter states, are now expected to vanish.
This means that vertices of the type 2) should now be suppressed
with respect to the diagrams of the type 1).
It is therefore reasonable to expect a change in the sign of the 
``beta-function''. We stress however that these arguments have to be
taken with extreme care, being only extremely qualitative.

We are now in the position to reconsider the puzzle of the apparent 
insensitivity of
the string gauge threshold corrections to the existence (i.e. masslessness)
of the hidden sector, in contrast with the dependence on their coupling.
It is now clear that we cannot anymore rely on ``index'' theorems
in order to compute such corrections: at a generic point in the 
K3 moduli space, ${\cal N}_4=2$ supersymmetry may not exist.
We are therefore not allowed to advocate supersymmetry in order to
exclude any dependence of the threshold corrections on the K3 moduli,
and promote the result of the computation performed at an orbifold point,
to a general result valid at any point of the K3.
Actually, the explicit results sofar obtained 
for the heterotic string, have been computed at the $Z_2$ orbifold
point \cite{kkpr}. This is a very special point in the 
hypermultiplets moduli,
and there is no surprise that precisely at that point the 
hidden sectors are massless. In fact, at this point the
hypermultiplet moduli are twisted; it is very likely that the
masslessness of the hidden sectors is related to this, as much as
the two Horava--Witten walls providing the two $E_8$ give rise to
massless states only when the eleventh coordinate is twisted.
If, away from the $Z_2$ orbifold point, supersymmetry is broken, nothing
prevents the moduli of K3 from contributing to the correction:
very likely, the mass of the states of the hidden sectors acts 
as a suppression factor for the dependence of the threshold corrections 
on the gauge couplings of these sectors.

The above arguments, relating the non-vanishing of the beta-function to the
breaking of supersymmetry, apply to the gauge coupling, but not to
couplings like that of the $R^2$ term. 
More precisely, they don't  apply to the ``modified'' $R^2$ term,
the gravitational amplitude considered in this work, and introduced in
Ref. \cite{gkp}. This amplitude is precisely defined by subtracting the
``interaction'' terms of the currents, accounting for true ``loop
effects'' of the theory. 
The large-$T$, -$U$ contribution of the ``hidden sectors'' to this amplitude
is therefore a ``tree level'' effect of the theory, that accounts for
the presence of these extra sectors, ``weighting'' like the 
perturbative sector, and entering on the same footing.
As we saw, the beta-function of this term is non-vanishing for
all the cases considered in sections \ref{rank48}--\ref{complete}.

\vspace{1.5cm}

\noindent

\section{\bf Conclusions}
\label{conclusions}

In this work we have considered type~II/heterotic/type~I
string--string duality for constructions with
perturbative ${\cal N}=4$ and ${\cal N}=2$ supersymmetry in four dimensions.
Our aim was to investigate some non-perturbative properties of string theory
by using as source of information the various dual string constructions. 
In all the  cases we considered,
there exists a region in the moduli space that corresponds to a weak
coupling regime at the same time in the type~I, type~II and heterotic strings.
For any example, we have therefore considered the theory from all the three
dual points of view: no one of these single approaches can in fact capture 
at once all the aspects of a construction. 
The existence of a common region of weak coupling 
does not mean that the theory has to look the same in all the 
three dual string approaches: the existence of non-perturbative states
present for any value of the string coupling makes in some cases
the identification of dual constructions and the test of duality quite subtle. 
In such cases, an analysis based on a naive comparison
of the moduli spaces can be misleading, being necessarily affected by the 
identification of the associated effective theories.
We therefore performed the analysis at the $Z_2$ orbifold point, where we can
explicitly write the partition function and compute string threshold
corrections.
The comparison of the renormalization of certain couplings of the effective 
action allows us to point out several novel features, that in certain cases
lead to an identification of duals 
slightly different from what previously proposed, thereby changing
the perspective of the duality map between string constructions.
Among the cases considered, particularly relevant are those in which
part of the string massless spectrum is non-perturbative from the heterotic 
and/or type~I point of view. These provide explicit examples of the
situations considered in Refs. \cite{48,poster}, of four dimensional 
heterotic theories obtained by toroidal compactification on $T^2$ of a six
dimensional theory, in which however the ${\cal N}_4=2$ heterotic 
non-perturbative massless spectrum does 
not fit with the anomaly constraint of ${\cal N}=1$ field theory 
in six dimensions. The investigation of these theories from
the dual type~II point of view makes clear that, 
although from the heterotic point of view the ${\cal N}_4=2$ theory is just
obtained by toroidal compactification of ${\cal N}_6=1$, in their whole
these theories cannot be defined in six dimensions, in the sense that the
massless states cannot be represented all at the same time in terms of 
an effective ${\cal N}_6=1$ field theory. The mismatch
is on the other hand in a part of the spectrum which is non-perturbative 
from the heterotic point of view: there is therefore no contradiction with
the field theory constraints, that control only the perturbative part 
of the heterotic string, the one corresponding to a six dimensional
supergravity description.
Indeed, in the case of ${\cal N}_4=2$ orbifolds,
we found an intriguing relation between \romannumeral1) the T-duality of the
heterotic string, that exchanges the K\"{a}hler class and complex structure
moduli, resp. $T$ and $U$, of the two torus, \romannumeral2)
the modular invariance on the type~II side, that implies a symmetry between 
different orbifold twisted sectors, and \romannumeral3) the type~I symmetry 
between D9- and D5-branes sectors. 
For string theory, modular invariance is a more fundamental
property than anomaly cancellation in the associated effective action;
only when one considers just the perturbative string spectrum the
two requirements in some cases turn out to be equivalent. 
The counter-example is provided by the above mentioned situations. 
There, only a part of the heterotic string spectrum satisfies $d=6$ anomaly
cancellation constraints, and admits a representation in terms of an effective
supergravity also when decompactified to six dimensions.
Nevertheless, the string theory is well defined; in particular,
the massless spectrum is consistent with pure string theory requirements,
such as modular invariance, or T-duality, or tadpoles cancellation.
These properties, intriguingly related each other and, ultimately, to the 
renormalizability of string theory, are
non-trivially interchanged under string--string duality.

Another pure stringy phenomenon, that does not have a field theory counterpart,
is the ``interaction'' of perturbative and ``hidden'', non-perturbative 
sectors, that takes place whenever in the heterotic, or type~I string,
some gauge group factors possess a non-vanishing beta-function.
The kind of interaction between sectors is rather peculiar and cannot be
written in terms of an effective field theory with multi-dilaton gauge sectors.
Indeed, the expressions of the threshold corrections suggest that in 
such cases the hidden sector is at the strong coupling. As a  consequence, 
supersymmetry is broken by gaugino condensation in the
hidden sectors. This scenario is compatible with
a perturbatively supersymmetric construction, because 
the supersymmetry breaking terms  are non-perturbative 
from the point of view of the ``visible'' sector. 
We only devoted some qualitative comments to this phenomenon, that
definitely deserves a deeper understanding and further investigation.
A consequence of this is in fact the general breaking to ${\cal N}=0$ in 
the heterotic string with ${\cal N}_4=1$.   

With this work, we hope at least to have made clear the urgency of looking 
at string theory by avoiding as much as possible any short cut
provided by a heavy use of the properties of its representations in terms of 
supersymmetric field theories, an approach sometimes misleading.

\newpage
\noindent

\vskip 1.5cm
\centerline{\bf Acknowledgments}
\noindent
I am grateful to Jean--Pierre Derendinger and to Robert Helling
for valuable discussions. I also acknowledge the kind hospitality
of the Department of Physics of the University of Neuch\^atel and the
Swiss National Science Foundation for partial financial support.

\vspace{1.5cm}

%\newpage

%\bibliography{robert}
%\bibliographystyle{ieert}

\providecommand{\href}[2]{#2}\begingroup\raggedright\endgroup

\end{document}